\documentclass[journal,10pt]{IEEEtran}
 \usepackage{multirow}
\usepackage{cite}
\usepackage{xcolor}

\usepackage{soul} % for \hl

\definecolor{lightblue}{rgb}{0.63, 0.79, 0.95}

\usepackage{booktabs}
\usepackage{amsthm}
\theoremstyle{plain}
\newtheorem{thm}{Theorem}

\usepackage{url}
\usepackage{breakurl}
\usepackage[breaklinks]{hyperref} % autoref
\ifCLASSOPTIONcompsoc
	\usepackage[caption=false,font=normalsize,labelfont=sf,textfont=sf]{subfig}
\else
	\usepackage[caption=false,font=footnotesize]{subfig}
\fi
\usepackage[pdftex]{graphicx}
\usepackage{enumitem}
\usepackage{amsfonts}
\usepackage{amsmath}
\newcommand*{\Comb}[2]{{}^{#1}C_{#2}}%
\usepackage{amssymb}
\usepackage{siunitx} % \degree
\usepackage{multirow}

\usepackage{epstopdf}

\begin{document}

\title{EPIC: Efficient Privacy-Preserving Scheme with E2E Data Integrity and Authenticity for AMI Networks}

\author{Ahmad~Alsharif,%~\IEEEmembership{Member,~IEEE,}
	~Mahmoud~Nabil,%~\IEEEmembership{Member,~IEEE,}
    ~Samet~Tonyali,%~\IEEEmembership{Member,~IEEE,}	
    ~Hawzhin~Mohammed,%~\IEEEmembership{Member,~IEEE,}
		~Mohamed~Mahmoud,~\IEEEmembership{Member,~IEEE,}
	~and~Kemal~Akkaya,~\IEEEmembership{Senior~Member,~IEEE,}
		\vspace{-6mm}

\thanks{A. Alsharif is with the Department of Computer Science, University of Central Arkansas, Conway, AR, 72035 USA and also with the department of Electrical \& Computer Engineering,  Tennessee Tech University, Cookeville, TN 38505 USA. %\protect\\
	E-mails:  aalsharif@uca.edu}% <-this % stops a space

\thanks{M. Nabil, H. Mohammed and M. Mahmoud are with the Department of Electrical \& Computer Engineering,  Tennessee Tech University, Cookeville, TN 38505 USA. %\protect\\
	E-mails:  mnmahmoud42@students.tntech.edu, hmohammed42@students.tntech.edu, and mmahmoud@tntech.edu}% <-this % stops a space

\thanks{S. Tonyali and K. Akkaya are with the Department of Electrical \& Computer Engineering, Florida International University, Miami, FL 31174 USA.
	Emails: stony002@fiu.edu, kakkaya@fiu.edu}

%\thanks{Manuscript received XXX, XX, 2017; revised XXX, XX, 2017.}
} % end of \author

%\markboth{IEEE Internet of Things Journal,~Vol.~XX,~No.~XX,~XXX~2018}
{}
%{Shell \MakeLowercase{\textit{et al.}}: Bare Demo of IEEEtran.cls for Journals}

\maketitle

\begin{abstract}

In Advanced Metering Infrastructure (AMI) networks, smart meters should send fine-grained power consumption readings to electric utilities to perform real-time monitoring and energy management.
However, these readings can leak sensitive information about consumers' activities.
Various privacy-preserving schemes for collecting fine-grained readings have been proposed for AMI networks.
These schemes aggregate individual readings and send an aggregated reading to the utility,
but they extensively use asymmetric-key cryptography which involves large computation/communication overhead.
Furthermore, they do not address End-to-End (E2E) data integrity, authenticity, and computing electricity bills based on dynamic prices.
In this paper, we propose EPIC, an efficient and privacy-preserving data collection scheme with E2E data integrity verification for AMI networks.
Using efficient cryptographic operations, each meter should send a masked reading to the utility such that all the masks are canceled after aggregating all meters' masked readings, and thus the utility can only obtain an aggregated reading to preserve consumers' privacy.
The utility can verify the aggregated reading integrity without accessing the individual readings to preserve privacy.
It can also identify the attackers and compute electricity bills efficiently by using the fine-grained readings without violating privacy.
Furthermore, EPIC can resist collusion attacks in which the utility colludes with a relay node to extract the meters' readings.
A formal proof, probabilistic analysis are used to evaluate the security of EPIC, and ns-3 is used to implement EPIC and evaluate the network performance.
In addition, we compare EPIC to existing data collection schemes in terms of overhead and security/privacy features.

\end{abstract}


\IEEEpeerreviewmaketitle

\vspace{-5mm}
%\IEEEraisesectionheading{\section{Introduction}\label{sec:introduction}}
\section{Introduction}\label{sec:introduction}

%%%%%\IEEEPARstart{T}{he} power grid is the biggest interconnected network on earth.
%%%%%%It can generate more than one million Megawatts of power by using more than 9,200 electric generating units worldwide \cite{PT01}.
%%%%%However, the current grid is unreliable, inefficient, and a major pollutant to the environment.
\IEEEPARstart{T}{he} current power grid is unreliable, inefficient, and a major pollutant to the environment.
Recent reports indicated that the power outages cost the United States (U.S.) at least 150 billion dollars each year \cite{HT05}.
Also, the north-east blackout in August 2003, which lasted for a week, affected over 100 power plants and about 55 million people \cite{newro1}.
Investigations showed that this blackout could be avoided if the grid could provide effective real-time diagnostic support \cite{newro}.
%%%%%Furthermore, according to the U.S. Energy Information Administration, the emissions of CO$_2$ by the U.S. electric power grid reached 1,925 million metric tons, or about 37\% of the total U.S. energy-related CO$_2$ emissions \cite{PT03}.

%In AMI network, fine-grained power consumption collection is required for real-time monitoring, energy management, and dynamic billing.However, fine-grained power consumption collection violates consumers' privacy, and the fine-grained readings could be modified in the way between the reporting meter and the utility.

The smart grid initiative aims to develop a clean, reliable, and efficient system.
It extensively integrates information and communication technologies into the power grid \cite{ICT_agg}.
One of the main components of the smart grid is the Advanced Metering Infrastructure (AMI) networks that connect smart meters (SMs) installed at consumers' side to the electric service provider (the utility).
SMs should send fine-grained power consumption data to the utility to perform real-time monitoring and energy management \cite{HT07}.
%reduction to the peak-to-average ratio which can help preventing brownouts, a reduction of available electricity in a particular area, and blackouts , a failure of supplying electricity \cite{HT07,report_june_2017,Fine_grained_app1}.
Moreover, the utility can reduce the power consumption during peak hours using dynamic pricing approach in which the electricity prices may change during the day to encourage consumers to reduce their power consumption.

However, the fine-grained power consumption readings can reveal sensitive information about the consumers' activities, such as the times consumers leave/return homes, as well as, the appliances they use since each appliance has a unique power consumption signature \cite{HT01,HT02,enhanced}.
%On the other hand, electric appliances such as refrigerators, air conditioning, etc, generate unique load signatures.
%Therefore, sending fine-grained power consumption can be used to violate consumers privacy by performing real-time surveillance, i.e. profiling consumers activities, determining which household appliances are being used, or learning times at which they sleep/leave/return home \cite{usage_graph,HT01,HT02}.
Privacy-preserving data aggregation is a promising technique to enable the utility to obtain an aggregated fine-grained reading from an AMI network without learning the individual readings to preserve the consumers' privacy.
However, the existing schemes, such as \cite{HT20,6165271,cube,li2018ppma,DI01}, extensively use asymmetric-key cryptography in data aggregation, which typically involves large computation and communication overhead.
They also do not address End-to-End (E2E) data integrity in which the utility can ensure that all the individual fine-grained readings are not altered during transmission and aggregation without accessing the individual readings to preserve privacy.
Moreover, they do not address E2E authenticity in which the utility can ensure that the aggregated reading is computed using the fine-grained readings coming from intended consumers.
Furthermore, generating electricity bills using the reported fine-grained readings based on dynamic prices is challenging
since the utility should not have access to the fine-grained readings to preserve privacy, but these readings are needed to generate consumers' bills.

%, and the ability to generate electricity bills based on dynamic prices efficiently by using the fine-grained readings while preserving consumers privacy.
In this paper, we propose an \textbf{E}fficient \textbf{P}rivacy-preserving scheme with E2E data \textbf{I}ntegrity, authenticity and \textbf{C}ollusion-resistance for AMI networks, named ``\textbf{EPIC}''.
The idea is that each SM selects a number of SMs in the network called ``\textit{proxies}'' and efficiently computes shared pairwise secret masks with each proxy.
Then, it should mask its fine-grained reading with all the masks shared with the proxies, such that all the masks are canceled after aggregating all meters' masked readings, and thus the utility can only obtain an aggregated reading to preserve consumers' privacy.
EPIC can also resist collusion attacks in which the utility can collude with a relay meter to extract a meter's fine-grained readings because readings are masked by several secret masks shared with a number of different proxies.
The number of the selected proxies controls the protection level against collusion attack.
In addition, to ensure E2E data integrity and authenticity, a homomorphic hash and a hash MAC are computed on each masked reading.
Then, hash MACs are aggregated while all the individual homomorphic hashes are forwarded to the utility.
Using the individual homomorphic hashes and the aggregated MAC, the utility can ensure the data integrity of each individual fine-grained reading and the authenticity of each consumer in the network.
Furthermore, the homomorphic hashes can be also used to enable the utility to generate dynamic electricity bills without accessing the individual readings to preserve privacy.

%To address the aforementioned problems, we propose in this paper an \textbf{E}fficient \textbf{P}rivacy-preserving scheme with E2E data \textbf{I}ntegrity verification and \textbf{C}ollusion-resistance for AMI networks named ``\textbf{EPIC}''.
%The idea is that, each meter selects a number of meters in the network called ``\textit{proxies}'' and efficiently computes and shares pairwise secret mask with each proxy.
%Then, each meter should mask its fine-grained reading with all the individual shared masks and send the utility a masked power consumption reading.
%The masks are computed such that all the masks are canceled after aggregating all meters' masked readings, and thus the utility can only obtain an aggregated reading to preserve consumers privacy.
%\hlgreen{Based on data masking} \cite{Hawzhin}, BLS short signature \cite{boneh2001short}, homomorphic hashes \cite{hom_hash}, and message authentication codes, EPIC can achieve collusion resistance, E2E data integrity, and dynamic billing simultaneously.

Our contributions can be summarized as follows.

\begin{itemize}
	\item \textit{Efficient and collusion-resistant privacy-preserving power consumption collection.}
	EPIC uses secure and lightweight operations to efficiently mask the fine-grained readings and aggregate the masked readings to enable the utility to collect a fine-grained aggregated reading without leaking the consumers' sensitive information.
	It can also resist collusion attacks to reveal a meter's readings and allows the SMs to set their protection level.

	%in EPIC, the utility needs to collude with all the meter's proxies to be able to extract a meter's reading, and thus the meters can control the level of protection against collusion attacks by selecting a proper number of proxies.

	\item \textit{E2E data integrity.}
	Since the meters' reading can be modified during the transmission to the utility, EPIC enables the utility to verify the integrity of the aggregated reading without accessing the individual readings to preserve consumers' privacy.	
	It also enables the utility to identify the attackers who modify the readings.
	
	\item \textit{E2E authenticity.}
	In EPIC, the utility can ensure that the aggregated reading was computed by the intended users in the network.

	\item \textit{Dynamic pricing.}
	Using homomorphic hash properties, EPIC enables the utility to efficiently compute electricity bills based on dynamic pricing without violating consumers' privacy.
\end{itemize}

The results of a formal proof, probabilistic modeling, and analysis demonstrate that EPIC is secure.
In addition, ns-3 is used to implement EPIC and evaluate the network performance.
The results demonstrate that our scheme is efficient.
We also compare EPIC to the existing data collection schemes in terms of overhead and security/privacy features.

%The performance of EPIC is evaluated using ns-3 and the measurements demonstrate that EPIC is much more efficient than the existing schemes. In addition, our analysis demonstrates that EPIC can preserve the consumers' privacy with end-to-end data integrity verification and achieving satisfactory protection against collusion attacks.

A preliminary version of this paper appeared in \cite{Hawzhin}.
The main difference between the conference version and the current version are as follows.
First, \cite{Hawzhin} did not address E2E data integrity, E2E authenticity, attacker identification, dynamic pricing, and details of key management and sharing secret masks offline and efficiently. The journal version addresses all these challenges.
Second, extensive analysis and simulation have been added to the journal version. This includes a formal security proof, a comprehensive security analysis, probabilistic analysis of the collusion attacks and the proposed defense method, and updated ns-3 simulation results against similar existing schemes.

The remainder of the paper is organized as follows.
Section \ref{sec:Related-Work} discusses the related works.
The system models and preliminaries are presented in section \ref{sec:System-model}.
The proposed masking and aggregation method is presented in section \ref{sec:aggregation}.
The details of EPIC are given in section \ref{sec:Proposed-Scheme}.
The security and privacy analysis is given in section \ref{sec:Security-and-privacy}, whereas the performance evaluation and experimental results are given in section \ref{sec:Performance-Evaluation}.
Finally, conclusions are drawn in section \ref{sec:Conclusion}.

\section{Related Works}\label{sec:Related-Work}

%Several schemes have been proposed that target efficient power consumption collection, privacy preservation and data integrity for AMI networks.

Cryptographic and non-cryptographic schemes have been proposed to collect power consumption readings in AMI networks with the preservation of consumer privacy.
Non-cryptographic schemes hide the real power consumption using a rechargeable battery that modifies consumption patterns \cite{PP24}.
However, this approach is costly because it requires installing batteries at consumer sides and regularly maintaining them.
%However, such a system requires regular maintenance to the installed batteries, may induce a cost depending on battery performance and dynamic utility pricing, and does not provide the utility with an accurate readings.
For cryptographic schemes, the existing schemes %in the literature
use anonymization or data aggregation to preserve privacy.

\emph{Anonymization.}
The anonymization schemes aim to hide the real identity of the senders of the power consumption readings.
In \cite{PP01}, messages should follow different routes to the utility every time the meter reports its reading to hide the sender of the messages.
%In \cite{tan2016pseudonym, PP22}, short-term pseudonyms are used to anonymize the data sender.
In \cite{PP22}, short-term pseudonyms are used to anonymize the data sender.
%However these schemes are susceptible to data modification and collusion attacks.
However, since readings are sent every short time, the utility can link the readings of each meter, which degrades consumer privacy.
%\hlgreen{Moreover, the fine-grained readings are sent with no encryption, an eavesdropper can breach users' privacy.}

\emph{Data aggregation.}
%%%%%Data aggregation technique preserves consumer privacy by hiding the individual readings.
%%%%%The idea is that the individual readings should be aggregated and only an aggregated reading for all the meters in the AMI network should be sent to the utility.
Comparing to anonymization and non-cryptographic techniques, data aggregation is a practical and cheap approach for achieving high level privacy preservation.
In the literature, data aggregation is used  in AMI and wireless sensor networks (WSNs) \cite{HT20,6165271,cube,li2018ppma,DI01,erkin2012private,am03,DI14}.
%\cite{HT20,6165271,DI01,PP06,am03,DI04,DI14}. %For instance, in Paillier cryptosystem \cite{paillier1999public} $E(m_{1})\times E(m_{2})=E(m_{1}+m_{2})$ where $m_{1}$ and $m_{2}$ are the message and $E(m)$ is the encryption of $m$.
%However, homomorphic encryption suffers from two main drawbacks.
%First, all the data owners should encrypt there message with the same key which also should be known by the utility for decryption so a trusted authority come into the scene.
%Secondly, the ciphertext size in homomorphic encryption is large for the same level of security compared to other symmetric key encryption methods, this may incurs more communication overhead.
%Despite these issues, homomorphic encryption is still very popular literature.
In \cite{HT20}, Fan et al. use bilinear pairing with an aggregation method based on blind factors and solving the discrete log problem using Pollard's lambda method
%%%%%\cite{boneh2005evaluating}
to obtain the aggregated reading and achieve collusion resistance.

In \cite{6165271}, Lu et al. used homomorphic encryption to aggregate multi-dimensional data represented using a superincreasing sequence.
Shen et al. in \cite{cube} proposed cube-data aggregation by using the Paillier cryptosystem and Horner’s Rule.
In \cite{li2018ppma}, Li et al. proposed the use of two superincreasing sequences with the Paillier cryptosystem to achieve multi-subset data aggregation.

In \cite{DI01}, Li et. al. used homomorphic-encryption-based aggregation scheme to send an aggregated reading to the utility.
The utility should run anomaly detection system in every data collection round to detect data modification attack, but unlike EPIC, the system suffers from false positive and negatives.

%%%%%Unlike traditional solutions that assume a single electric supply model, Mustafa et. al. propose a multi-entity model that uses homomorphic encryption and short signatures in \cite{PP06}.
In \cite{am03} Garcia et al. combined Paillier's homomorphic encryption with additive secret sharing scheme to protect the scheme against collusion attacks.
However, the encryption and aggregation complexities of \cite{am03} are $\mathcal{O}(n)$ and $\mathcal{O}(n^2)$ respectively while in EPIC they are $\mathcal{O}(1)$ and $\mathcal{O}(n)$.
%In \cite{DI04}, Ozdemisr et. al. used homomorphic encryption with message authentication codes to provide secure and privacy preserving scheme for WSNs.
In \cite{DI14}, Li et. al. used homomorphic MAC and homomorphic hash functions to provide data integrity in WSNs against external attackers only with the assumption that all internal nodes are trusted.
Compared to EPIC, homomorphic encryption based aggregation schemes such as \cite{6165271,DI01,cube,li2018ppma,erkin2012private,am03} are inefficient as they require much larger size ciphertext and much more time for encryption, decryption, and aggregation than EPIC.

Different from homomorphic-encryption-based aggregation schemes, secure multi-party computation (SMC) based aggregation schemes were developed to achieve privacy through aggregation.
%another scheme named \emph{Secure Multi-party Computation} (SMC) is also used in literature.
SMC was introduced first in \cite{yao1982protocols} where each node splits its data into $k$ blocks such that the sum of all $k$ blocks is equal to the nodes' value.
Then, it randomly selects $k-1$ other nodes and sends to each of them a distinctive block.
The receiving nodes should aggregate the blocks it receives and transmits the result to the next node and so on.
%SMC-based aggregation is used in \cite{PP30} and the proposed scheme was extended to address  collusion attacks in \cite{PP31}.
In \cite{danezis2013smart}, Danezis et al. use SMC for computing non-linear functions such as multiplication, mean and variance for smart grid.
Conventionally, SMC incurs high communication overhead and transmitting shares to all other network nodes which affect the scheme scalability.

In \cite{mask_ring}, Knirsch et al. proposed the use of one-time masking for privacy-preserving data aggregation.
Specifically, in each data collection round, the SMs are arranged in a ring-based topology to sequentially update the smart meter masks before masking each fine-grained reading.
However, the proposed scheme has several limitations.
First, \cite{mask_ring} can only support single-hop model with a ring-topology communication used for online masks agreement, while EPIC can support both single- and multi-hop models with efficient and offline mask agreement as will be explained later in \autoref{sec:System-model} and \autoref{sec:aggregation}.
Second, in each data collection round in \cite{mask_ring}, all the ring SMs must communicate sequentially to ensure the correctness of masks updates before the actual reading reporting to the utility begins. This, therefore, increases the time required by the utility to collect the fine-grained readings and limits the network scalability.

%He et al. \cite{he2007pda} used the collaboration aggregation by proposing two schemes for wireless sensor networks.
%The first scheme is Cluster-based Private Data Aggregation (CPDA) where the sensors construct clusters randomly and they cooperate to compute the cluster aggregation result.
%In SMART, each node splits its data into $k$ slices and sends $k-1$ slices to nearest nodes via secure channel.
%However, both schemes suffer from high communication overhead.
%In \cite{DI06} He et al. improved their scheme by providing the data integrity.

%In \cite{PP07}, Haddad et. al.  utilized the bilinear pairing properties with data masking to achieve essential security requirements in AMI such as authentication, confidentiality, key agreement, and data integrity without trusting the Long Term Evolution Advanced (LTE-A).

\begin{table}[!t]

	\caption{\label{tab:Comparisons}Comparison between the proposed scheme and
		related schemes. }
	
	\center \scalebox{0.9}{
		
		{}%
		% Increase Row height
		\renewcommand{\arraystretch}{1.15}
		\begin{tabular}{lcccccc}
			\toprule
			& {EPIC}                     & {\cite{HT20}} & {\cite{6165271}}   & {\cite{cube}} & {\cite{li2018ppma}}   & {\cite{DI01}} \tabularnewline
			\midrule
			{Privacy Preservation}       & {$\surd$}     & {$\surd$}  & {$\surd$}  & {$\surd$}  & {$\surd$}  & {$\surd$}  \tabularnewline
			\midrule
			{Efficient Aggregation}      & {$\surd$}     & {NA}      & {NA}  & {NA}   & {NA} & {NA}       \tabularnewline
			\midrule
			{E2E Data Integrity  \&} & \multirow{2}{*}{$\surd$} & \multirow{2}{*}{NA} & \multirow{2}{*}{NA}  & \multirow{2}{*}{NA} & \multirow{2}{*}{NA} & \multirow{2}{*}{NA$^*$} \\
		{Attacker Identification}   &     &     &     &   &   &      \\ \midrule

{Collusion Resistance}       & {$\surd$}     & {$\surd$} & {NA}  & {NA}  & {$\surd$} & {NA} \tabularnewline
			\midrule
			%{E2E Data Integrity \& Attacker Identification} & {$\surd$}     & {NA}      & {NA}  & {NA$^*$} & {NA} \tabularnewline
			%\midrule

			{Support Dynamic Pricing}    & {$\surd$}     & {NA}      & {NA}  & {NA} & {NA} & {NA}  \tabularnewline
			\bottomrule
			\tabularnewline
		\end{tabular}{}}\\
	NA: Not Addressed\\
	$^*$: Unlike EIPC that uses cryptography to detect data modification, \cite{DI01} depends on anomaly detection to verify the data, which is vulnerable to false positives and negatives. \vspace{-5mm}
\color{black}
\end{table}

Homomorphic linear authenticators (HLAs) \cite{HLA} have been widely used to achieve data integrity for cloud applications \cite{HLA_based1,HLA_based2}. In cloud-based applications, each user breaks its data into several blocks, uses its private key to generate an authentication tag for each block and stores the data blocks and authentication tags on a cloud server. For data retrieval, a verifier sends a random challenge to the server and then uses the server response along with the users' public keys to ensure data integrity. Therefore, in cloud-based applications, data is not relayed by other users, instead, the data modification attacks can be launched by the cloud server.
Different from HLA-based schemes \cite{HLA_based1,HLA_based2}, EPIC considers a different network and threat models in which data is relayed by other SMs in the AMI network who may launch data modification attacks. Unlike \cite{HLA_based1,HLA_based2}, EPIC ensures data integrity by using the lightweight homomorphic hash along with an aggregated hash MAC to ensure data integrity as will be explained in \autoref{sub:utility_operations} and \autoref{sec:Attacks against data integrity}.

To sum up, we compare in Table \ref{tab:Comparisons} EPIC against similar schemes.
%in terms of the following features; privacy preservation, efficient aggregation, collusion resistance, E2E data integrity and attacker identification, and the support of dynamic pricing.
To the best of our knowledge, %and according to a recent survey \cite{HT30},
EPIC is the first solution that aims to achieve efficiency, privacy preservation, hop-by-hop and E2E data integrity and attackers identification, high resistance to collusion attacks, and dynamic pricing based billing simultaneously for both single- and multi-hop network models.

%To the best of our knowledge and from the survey that was conducted recently, 2016 \cite{HT30}, out of more than 170 papers on privacy-preservation and data integrity in smart grid communications, this is the first work in privacy preservation to use symmetric keys to preserve privacy. This is also the first work that combines privacy preservation, data integrity and collusion attacks.

\section{System Models and Preliminaries}\label{sec:System-model}

\subsection{Network Model} \label{sec:Network model} \vspace{0mm}

%\begin{figure}[!t]
%\centering \includegraphics[scale=0.5]{single-hop}
%\centering \includegraphics[scale=0.5]{multi-hop}
%%\centering \includegraphics[scale=0.5]{Structured}
%\caption{The considered system Model.}
%\label{fig:Network_Model}
%\vspace{-5mm}
%\end{figure}

\begin{figure}[!t]
	\centering
	\subfloat[Single-hop model. \label{fig:single-hop}]
	{\includegraphics[width=0.56\linewidth]{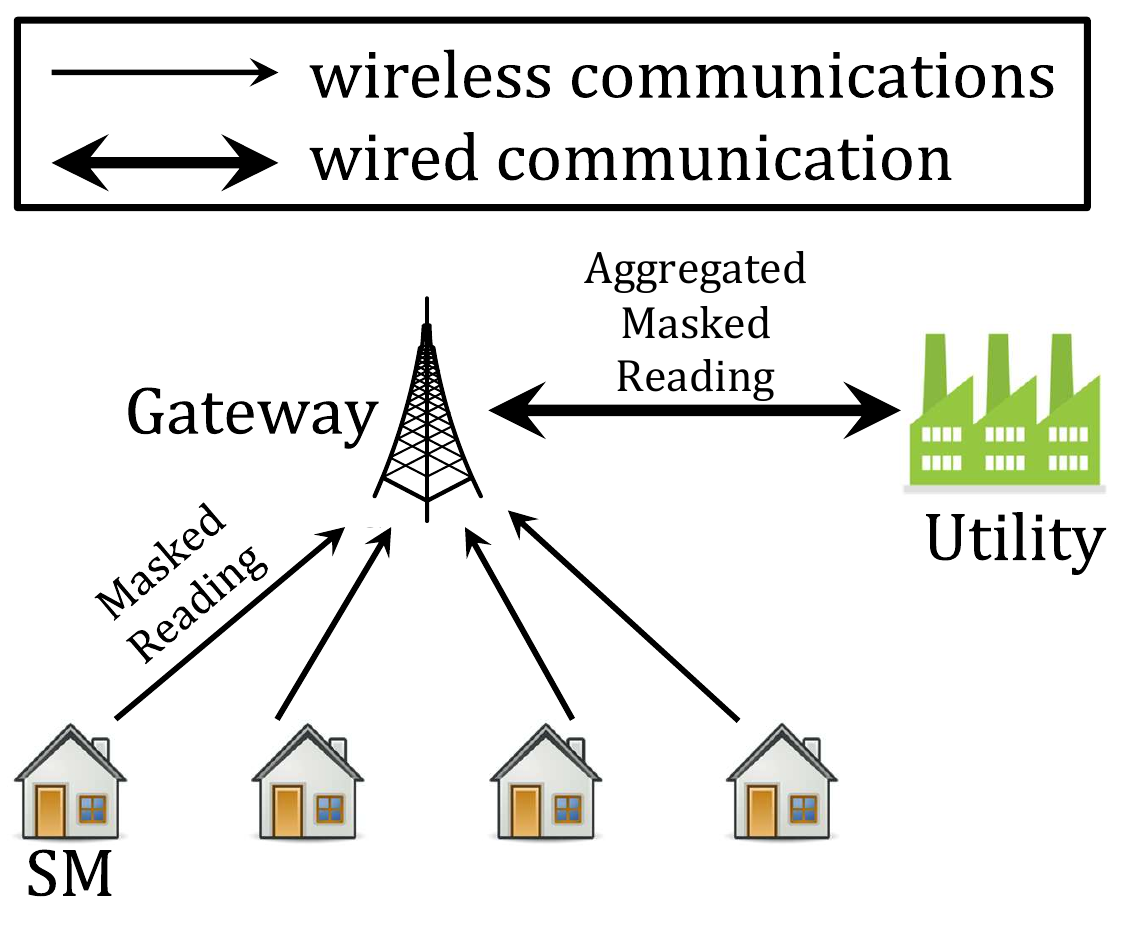}}
	
	\subfloat[Multi-hop model. \label{fig:multi-hop}]
	{\includegraphics[width=0.65\linewidth]{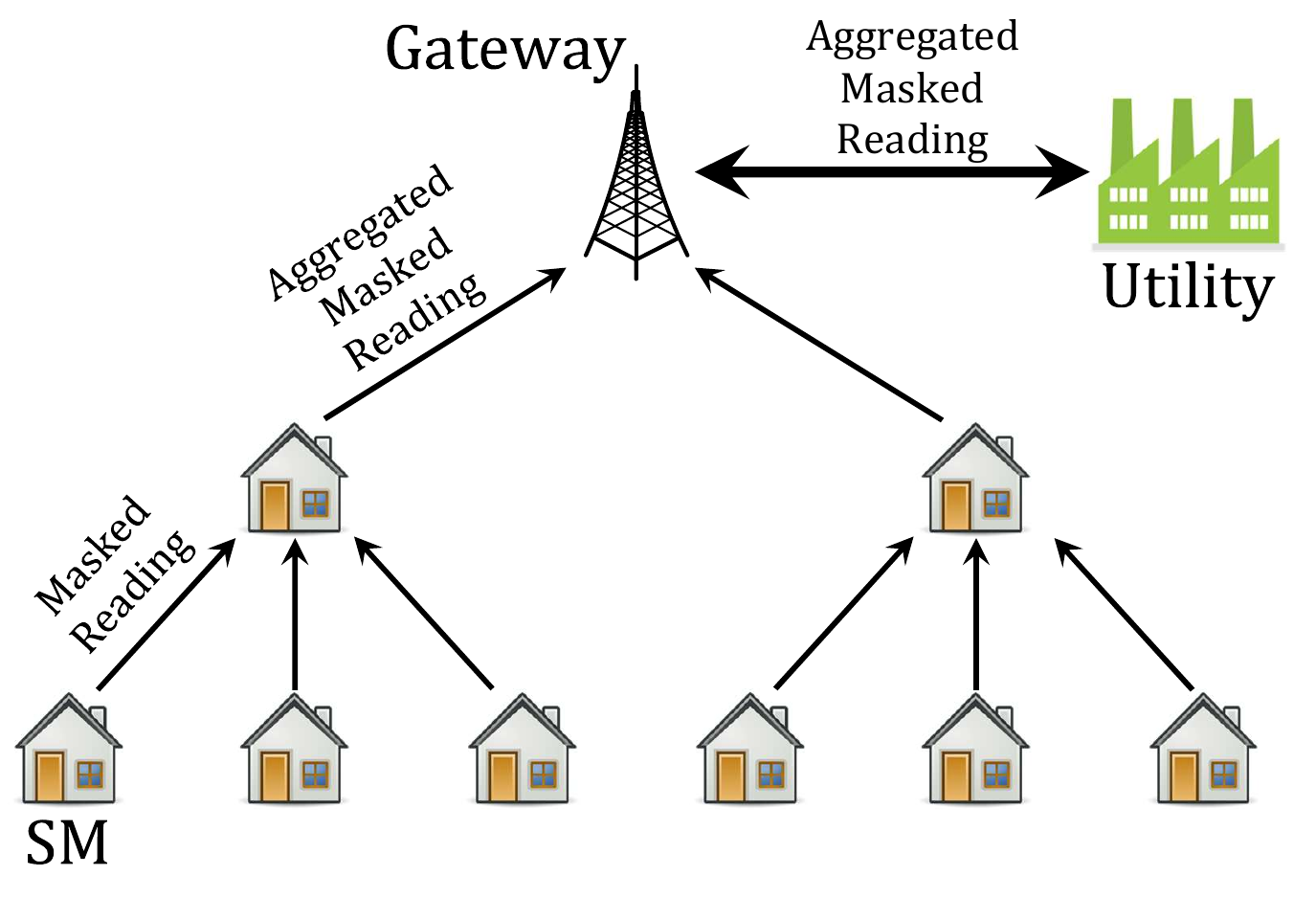}}	
	\caption{The considered network models.}
	\label{fig:Network_Model}
	\vspace{-6mm}
\end{figure}

As shown in Fig. \ref{fig:Network_Model}, the considered network model consists of the utility  and service subscribers in a residential area.
Each subscriber's house is equipped with a SM to report real-time fine-grained
power consumption readings to the utility every short time interval.
SMs can communicate with the utility through a local collector, called gateway.
As shown in the figure, EPIC can be used in single-hop or multi-hop network models.
The SMs are connected via a wireless mesh network using Wi-Fi where each meter can act as a router to relay meters' packets to connect them to the gateway.
The gateway can communicate to the utility through a wired link with low delay and high bandwidth.
For the single-hop model, SMs send reading packets to the gateway which aggregates all the readings, create a new packet, and send it to the utility.
For the multi-hop model, a virtual minimum spanning tree network topology that allows bottom-up aggregation is built. %\cite{PP14}.
Then, leaf SMs send their readings packets to their parent SM which uses its reading and the packets received from children SMs to create a new packet and forwards it to the next parent SM.
This should continue until the utility receives a reading packet.
%WiMax or LTE.

\subsection{Adversary Model}

\label{sec:Adversary and threat model} \vspace{0mm}

%To protect consumer's privacy, the meter masks and aggregates the readings.
%%%%%%We consider a strong adversary model, in which the
Attackers could be external adversaries $\mathcal{A}$, or internal network nodes, such as SMs, the gateway, or the utility.
Attackers may attempt to invade the consumers' privacy to learn their power consumption patterns.
They may also try to breach the data integrity by modifying other meters' data.
In addition, $\mathcal{A}$ can eavesdrop on all the communications between the different parties to infer any sensitive information about consumers.
$\mathcal{A}$ can also launch some active attacks such as packet replay and impersonation.
Moreover, the attackers can work individually or collude to launch stronger attacks.

\subsection{Preliminaries} \label{sec:preliminaries}

\subsubsection{Bilinear Pairing} \label{sub:pairing}
Let $\mathbb{G}_{1}$ be an additive cyclic group, $\mathbb{G}_{2}$ be a multiplicative cyclic group of the same prime order $q$, and $P$ be a generator of $\mathbb{G}_{1}$.
A pairing $\hat{e}:\mathbb{G}_{1}\times\mathbb{G}_{1}\rightarrow\mathbb{G}_{2}$ has the following properties.

\begin{itemize}
	%\vspace{1 mm}
	\item \textit{Bilinearity}: $\hat{e}(aP,bQ)=\hat{e}(P,abQ)=\hat{e}(abP,Q)=\hat{e}(P,Q)^{ab}\in\mathbb{G}_{2}$
	$\forall$ $P,Q\in\mathbb{G}_{1}$ and $a,b\in\mathbb{Z}_{q}^{*}$.
	%\vspace{1 mm}
	\item \textit{Non-degeneracy}: $\hat{e}(P,P)\neq1_{\mathbb{G}_{2}}$.
	%\vspace{1 mm}
	\item \textit{Computability}: $\hat{e}(P,Q)$ is efficiently computable $\forall$
	$P,Q\in\mathbb{G}_{1}$.
\end{itemize}

\subsubsection{Homomorphic Hash Function} \label{sub:homo_hash}
% $d=8$ as in \cite{hom_hash}
Let $\mathbb{G}$ be an additive cyclic group of prime order $p$ and has $d$ random generators $\{P_{1},P_{2},\dots,P_{d}\}\in\mathbb{G}$.
A homomorphic hashing on message $m=\{m_{1},m_{2},\dots,m_{d}\}$ can be constructed as
\begin{equation*}
\mathcal{H}(m)\stackrel{\text{def}}{=}\sum_{i=1}^{d}{m_{i}}P_{i}
\end{equation*}
Homomorphic hash function is collision resistant, where it is infeasible to find $m_1$ and $m_2$ such that $\mathcal{H}(m_{1}) = \mathcal{H}(m_{2})$.
In addition, homomorphic hash function is one way, where given $\mathcal{H}(m_{1})$, it is infeasible to compute $m_1$.
Homomorphic hash function also has the following property: $\mathcal{H}(m_{1}+m_{2})=\mathcal{H}(m_{1})+\mathcal{H}(m_{2})$.
We refer to reference \cite{hom_hash} for more details on homomorphic hash functions.

\begin{table}[!t]
	\center
	\caption{Main Notations.}
	\label{tab:Symbols-and-Abbreviations}
	\scalebox{0.95}{
		\begin{tabular}{cl}
			\toprule
			Symbol  & Meaning\tabularnewline
			\midrule
			SM$_{i}$  & Smart Meter $i$\tabularnewline
			\\[-0.9em]
			SM$_{pi}$  & Parent of Smart Meter $i$\tabularnewline
			$\ell_{i}$  & Total number of meters in SM$_{i}$ subtree\tabularnewline
			\\[-0.9em]
			$n_{i}$  & Number of direct children of SM$_{i}$\tabularnewline
			\\[-0.9em]
			$c$ & Children of SM$_{i}$ ($1\leqslant c\leqslant n_{i}$)\tabularnewline
			\\[-0.9em]
			$\sigma_{i}$  & Signature generated by SM$_{i}$\tabularnewline
			\\[-0.9em]
			$r_{i}$  & Fine-grained reading of SM$_{i}$\tabularnewline
			\\[-0.9em]
			$s_{i,j}$  & Secret mask shared between SM$_{i}$ and $\mathcal{P}_{i,j}$\tabularnewline
			\\[-0.9em]
			$m_{i}$  & Masked reading of SM$_{i}$\tabularnewline
			\\[-0.9em]
			$M_{i}$  & Aggregated masked readings for SM$_{i}$ subtree\tabularnewline
			\\[-0.9em]
			%$R$  & The total readings in the residential area \tabularnewline
			%\\[-0.9em]
			$h_{i}$  & Homomorphic hash on masked reading of SM$_{i}$\tabularnewline
			\\[-0.9em]
			$x_i$ ,$Y_i$  & private and public key of SM$_{i}$ \tabularnewline
			\\[-0.9em]
			$K_{i,j}$  & Symmetric key between node $i$ and node $j$\tabularnewline
			\\[-0.9em]
			HMAC$_{K_{i,u}}\left(h\right)$  & A keyed hash function on $h$\tabularnewline
			\\[-0.9em]
			$mac_{i,u}$  & HMAC $_{K_{i,u}}\left(h_{i}\right)$ \tabularnewline
			\\[-0.9em]
			$MAC_{i}$  & Aggregated MAC computed by SM$_{i}$\tabularnewline
			\\[-0.9em]
			$\mathcal{P}_{i,j}$  & Proxy $j$ of SM$_{i}$\tabularnewline
			\\[-0.9em]
			$\alpha_{i}$  & Number of proxies selected by SM$_{i}$ \tabularnewline
			\\[-0.9em]
			$\beta_{i}$  & Number of nodes that selected SM$_{i}$ as a proxy\tabularnewline
			\\[-0.9em]
			$\lambda_{i}$  & Total number of SM$_{i}$ proxies, $\lambda_{i}=\alpha_{i}+\beta_{i}$ \tabularnewline
			\\[-0.9em]
			%$TS$  & Time Stamp\tabularnewline
			%\\[-0.9em]
			%$u$  & Utility\tabularnewline
			%\\[-0.9em]
			%$gw$  & Gateway\tabularnewline
			%\\[-0.9em]
			$q,\mathbb{G}_{1},\mathbb{G}_{2},P,\hat{e}$ & Public parameters of bilinear pairing\tabularnewline
			\\[-0.9em]
			$\left\{ P_{1},...,P_{d}\right\} $,$\mathbb{G}$,$\mathcal{H}$  & Public parameters of homomorphic hash\tabularnewline
			\\[-0.9em]
			%$d$ & Number of homomorphic hash generators\tabularnewline
			%\\[-0.9em]
			%$E_K(.)$ , $D_K(.)$ & Symmetric key encryption/decryption schemes\tabularnewline
			\bottomrule
		\end{tabular}
	}
	\vspace{-2mm}
\end{table}

\section{Efficient and Collusion-Resistant Aggregation}\label{sec:aggregation}
In this section, we present a collusion-resistant and efficient data aggregation technique that is used in EPIC.
We refer to Table \ref{tab:Symbols-and-Abbreviations} for the main notations and parameters that will be used in this paper.

\subsection{Data Masking} \label{sub:Masking}

\begin{figure}[!t]
	\centering \includegraphics[scale=0.35]{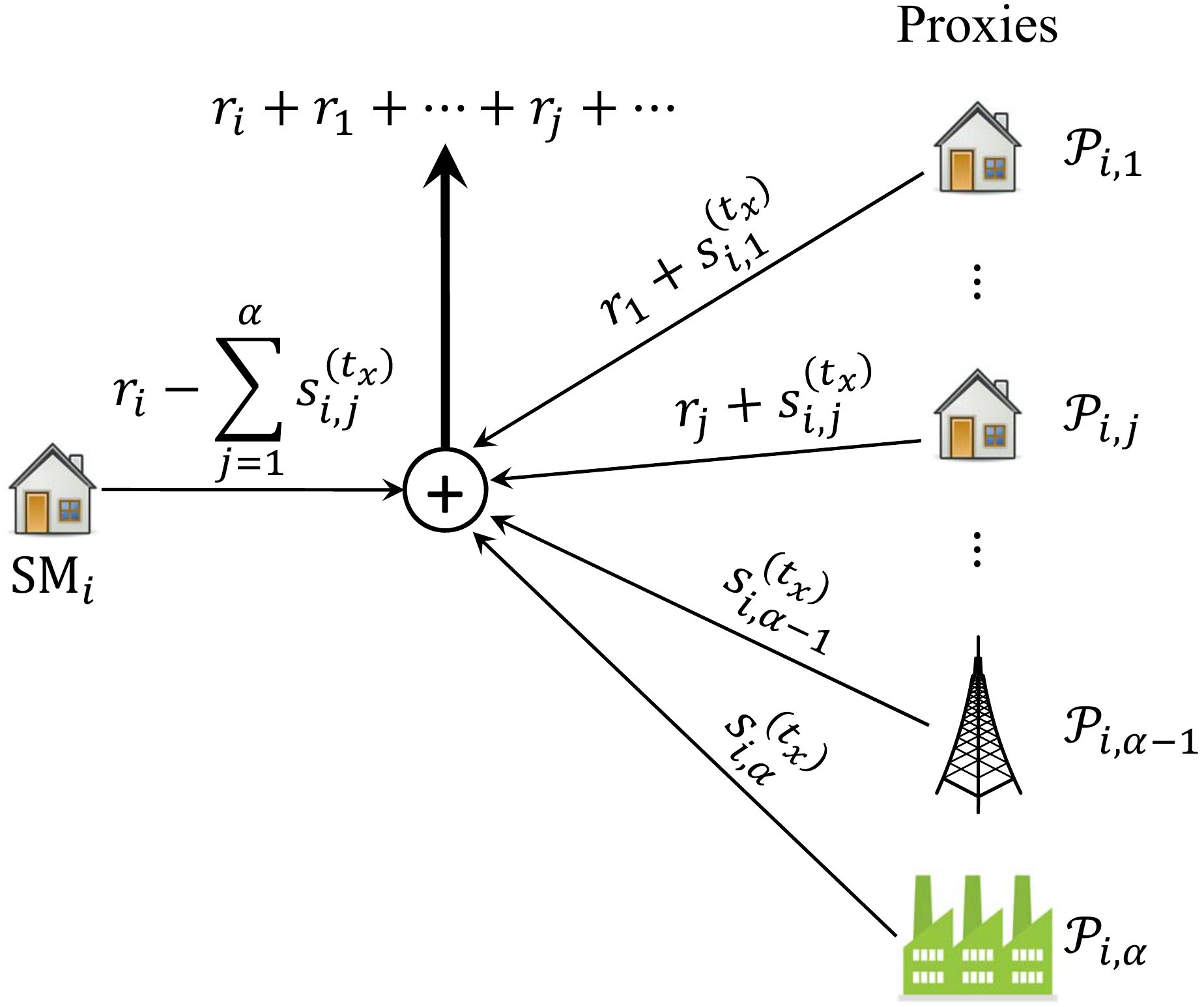}
	\caption{SM$_{i}$ masks its reading with secrets shared with proxies, and each proxy removes one secret from the mask by adding the mask to its reading.}
	\label{fig:mask}
	\vspace{-1mm}
\end{figure}

\emph{Masked Readings.}
Fig. \ref{fig:mask} illustrates the approach we use to protect consumer privacy and resist collusion attacks by using secret masks.
First, SM$_{i}$ chooses $\alpha$ proxies
%$\{\mathcal{P}_{i,1},\dots,\mathcal{P}_{i,j},\dots,\mathcal{P}_{i,\alpha}\}$
$\{\mathcal{P}_{i,1},\dots,\mathcal{P}_{i,\alpha}\}$ and shares a secret mask value $s_{i,j}^{(t_{x})}$ with each proxy $\mathcal{P}_{i,j}$ to be used for reporting the reading of time slot $t_x$.
As shown in the figure, each proxy $\mathcal{P}_{i,j}$ should add the mask $s_{i,j}^{(t_{x})}$ to its fine-grained reading $r_j$, whereas SM$_{i}$ masks its fine-grained reading $r_{i}$ by subtracting the summation of all shared masks with its proxies.
After aggregating all masked readings sent by SM$_{i}$ and its proxies, the masks added by all proxies cancel the mask used by SM$_i$.
\textit{If all the SMs follow this masking technique, the resultant value after the aggregation should be the summation of all fine-grained readings.}
Masks are used to achieve privacy preservation and collusion resistance as will be explained in section \ref{sec:Security-and-privacy}.
%It should be noted that all the proxies must collude in order to reveal the fine-grained reading of SM$_i$, and thus as the number of proxies increases, as the scheme becomes more resilient to collusion attacks because more proxies need to collude for a successful attack.

\emph{Mask Calculation.}
Each mask shared between a SM and each proxy should be used one time so that the masked readings look different even if the same reading is reported multiple time.
In addition, without changing the mask, the subtraction of two consecutive masked readings of a leaf SM can reveal the change in the power consumption by that SM.
%and changed in every report period otherwise two consecutive masked readings may reveal some information such as the change in the power consumption.
In EPIC, masks can be computed \textit{offline} and \textit{efficiently} as follows:
$s_{i,j}^{(t_{x})}=\text{HMAC}_{K_{i,j}^{(s)}}(Y_{i}$, $Y_{j}$, $day$, $t_{x})$,
where $\text{HMAC}_{K_{i,j}^{s}}()$ is a keyed hash function and ${K_{i,j}^{(s)}}$ is a short-time symmetric key that can be computed by the procedure that is explained in the next subsection, $Y_{i}$ and $Y_{j}$ are the public keys of the meter SM$_i$ and the proxy $\mathcal{P}_{i,j}$  respectively, $day$ is a unique day's date, and $t_x$ is a sequence number of the readings of one day.
Obviously, no other entity can derive the masks because it does not know the shared key.
%In addition, larger mask sizes can be generated by concatenating multiple hash values as\\
%$\text{HMAC}_{K_{i,j}^{(t_x)}}(Y_{i}$$\parallel$$Y_{j}$$\parallel$$t_{x}$$\parallel$$1)\parallel\text{HMAC}_{K_{i,j}^{(t_x)}}(Y_{i}$$\parallel$$Y_{j}$$\parallel$$t_{x}$$\parallel$$2)$$\parallel$$\dots$

\subsection{Efficient Key Agreement Procedure} \label{sec:Key agreemen}

Each meter needs to share a key with each proxy to efficiently calculate the secret mask.
%Also, each meter needs to share a key with each child meter and parent meter to secure the communication between the meters.
Key renewal is a good practice to thwart cryptanalysis attacks.
In this subsection, we describe two key agreement procedures to establish long-term and short-term keys.
Initially, a long-term symmetric seed key $K_{i,j}$, shared between SM$_i$ and its proxy $\mathcal{P}_{i,j}$ should be established and refreshed over a long period. %to reduce the network communication overhead.
Then, a short-term key $K_{i,j}^{(s)}$ is efficiently computed using the long-term seed key.

\subsubsection{Long-term seed key agreement}\label{sec:Infrastructure based AMI network}
\begin{figure}[!t]
	\centering \includegraphics[scale=0.45]{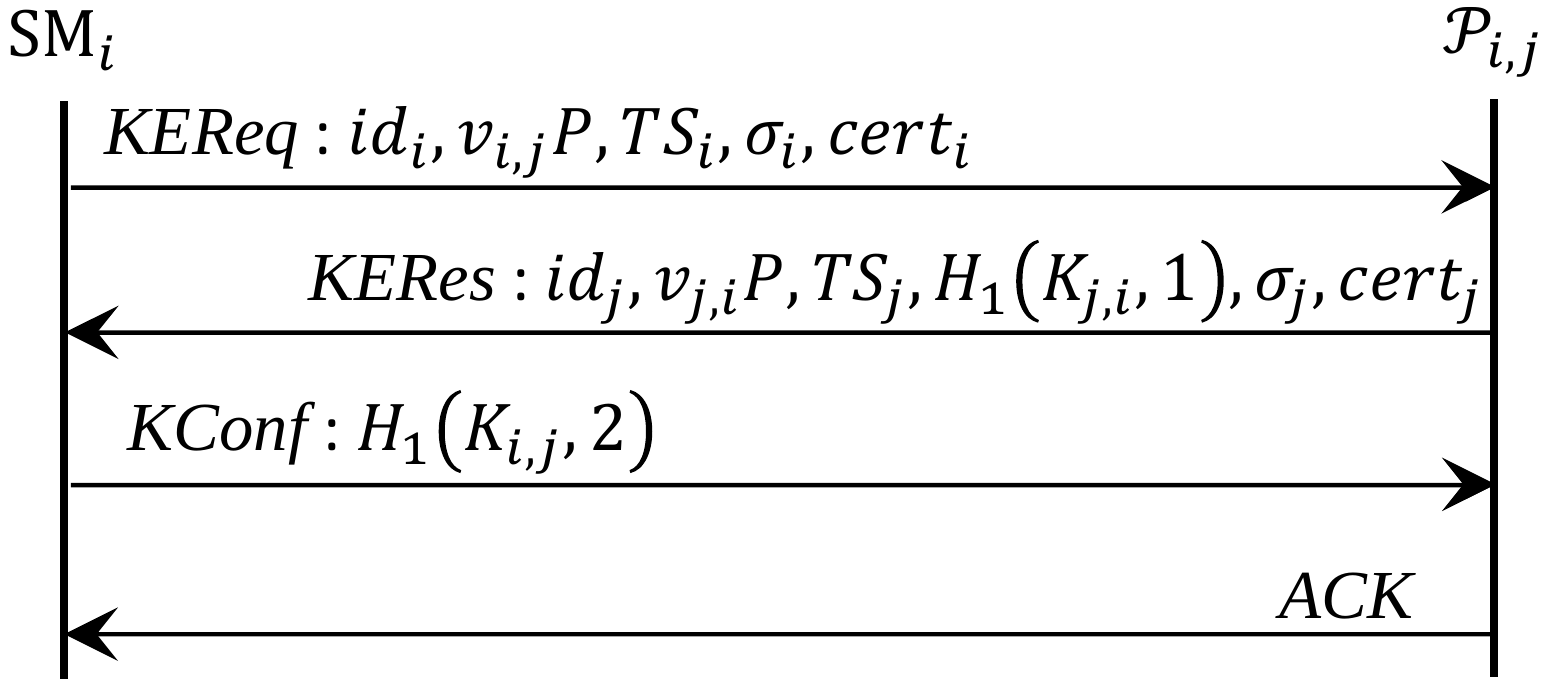}
	\vspace{-2mm}
	\caption{Long-term key establishment procedure.}
	\label{fig:KeyAgreement}
	\vspace{-5mm}
\end{figure}

To share a long-term seed key $K_{i,j}$ with each proxy $\mathcal{P}_{i,j}$, SM$_{i}$ chooses a random element $v_{i,j}\in\mathbb{Z}_{q}^{*}$ and composes  a key establishment request (\textit{KEReq}) packet as shown in Fig. \ref{fig:KeyAgreement}.
The packet contains $id_{i}$, $v_{i,j}P$, $TS_{i}$, $\sigma_{i}$ and $cert_{i}$, where, $id_{i}$ is the identifier of SM$_{i}$, $TS_{i}$ is the current timestamp, $v_{i,j}P$ is the random element $v_{i,j}$ multiplied by the generator $P$ of the additive group $\mathbb{G}_{1}$, $\sigma_{i}$ is a signature and $\sigma_{i}=x_{i}H_{2}\left(id_{i}, v_{i,j}P, TS_{i}\right)$, $H_{2}$ is a hash function defined as $H_{2}$:$\left\{0,1\right\}^{*}\rightarrow\mathbb{G}_{1}$, and $cert_{i}$ is the certificate of SM$_{i}$.
Finally, SM$_i$ sends the \textit{KEReq} packet to its proxies.
Each proxy $\mathcal{P}_{i,j}$ verifies that the packet is not stale by checking the timestamp $(TS_i)$ to thwart replay attacks.
Then, $\mathcal{P}_{i,j}$ uses SM$_i$'s public key, $Y_i=x_iP$, to verify the signature $\sigma_{i}$ by checking $\hat{e}\left(\sigma_{i},P\right)\stackrel{?}{=}\hat{e}\left(H_{2}\left(id_{i}, v_{i,j}P, TS_{i}\right),Y_{i}\right)$.
The signature verification proof is as follows:
\begin{eqnarray*}
	&  & \hat{e}\left(\sigma_{i},P\right)=\hat{e}\left(x_{i}H_{2}(id_{i}, v_{i,j}P, TS_{i}),P\right)\\
	&  & \hspace{12mm}=\hat{e}\left(H_{2}\left(id_{i}, v_{i,j}P, TS_{i}\right),x_{i}P\right)\\
	&  & \hspace{12mm}=\hat{e}\left(H_{2}\left(id_{i}, v_{i,j}P, TS_{i}\right),Y_{i})\right)
\end{eqnarray*}

If the signature is successfully verified, each proxy $\mathcal{P}_{i,j}$ chooses a random element $v_{j,i}\in\mathbb{Z}_{q}^{*}$ and computes $v_{j,i}P$.
Moreover, $\mathcal{P}_{i,j}$ calculates the long-term seed key $K_{j,i}=v_{j,i}v_{i,j}P$.
Finally, it sends the key establishment response (\textit{KERes}) packet to SM$_{i}$.
As shown in the figure, the packet has $id_{j}$, $v_{j,i}P$, $TS_{j}$, $H_1(K_{j,i}, 1)$, $\sigma_{j}$, and $cert_{j}$, where $\sigma_{j}=x_{j}H_{2}(id_{j}, v_{j,i}P, TS_{j}, H_1(K_{j,i}, 1))$, $H_1()$ is a hash function (such as SHA-1), and $H_1(K_{j,i}, 1)$ is used for key confirmation.
When SM$_{i}$ receives the packet, it checks the timestamp and verifies the signature similar to the verification process done by the proxy.
Then, it computes the long-term seed key $K_{i,j}=v_{i,j}v_{j,i}P=K_{j,i}$.
Finally, it sends a key confirmation packet (\textit{KConf}) to $\mathcal{P}_{i,j}$ so that $\mathcal{P}_{i,j}$ knows that SM$_i$ has successfully computed the long-term key.
%Finally, $\mathcal{P}_{i,j}$ responds with an acknowledgment message (\textit{ACK}) to acknowledge SM$_{i}$ that both of them have successfully shared the key and end the procedure.

\subsubsection{Short-term key computation}

\begin{figure}[!t]
	\centering \includegraphics[width=1\linewidth]{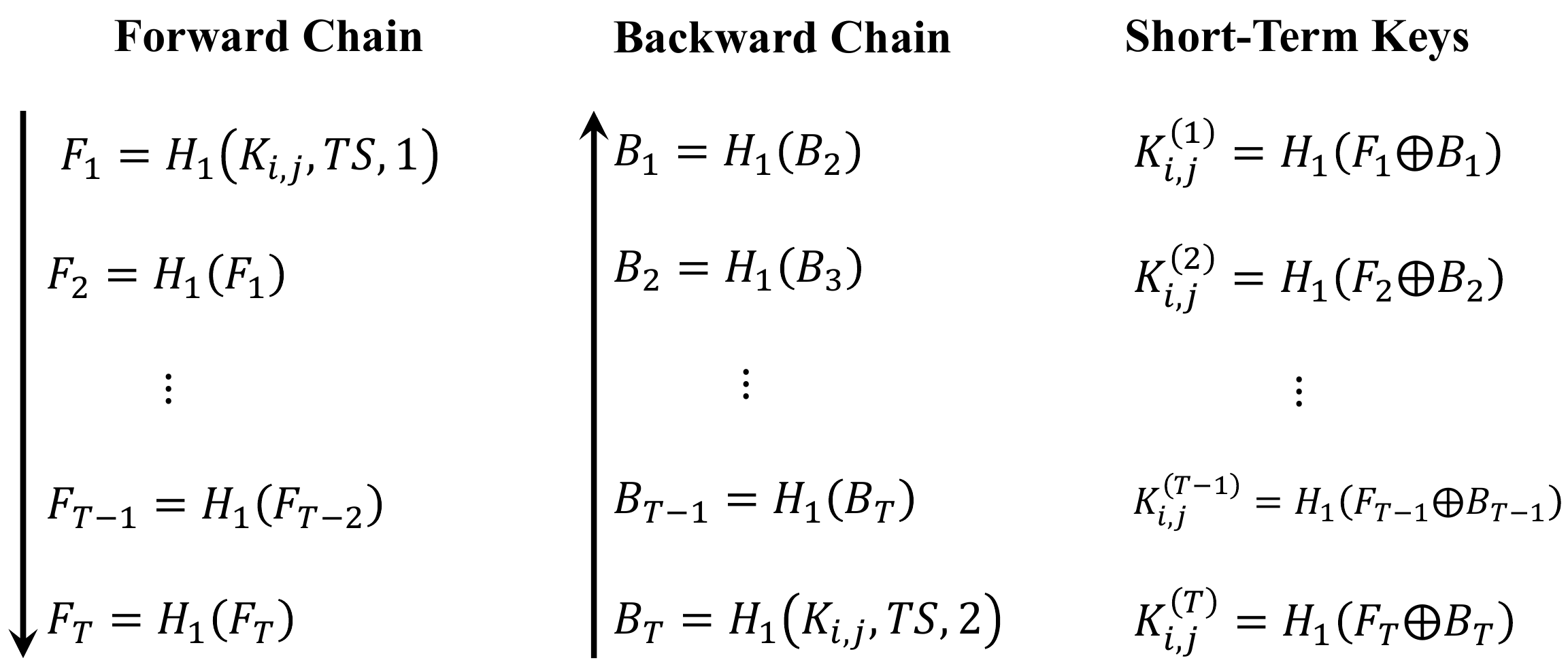}
	\caption{Short-term keys computation.}
	\label{fig:SessionKey}
	\vspace{-7mm}
\end{figure}

The long-term key $K_{i,j}$ is used as a seed for the computation of the short-term keys.
First, SM$_{i}$ and $\mathcal{P}_{i,j}$ use the seed key to compute bi-directional (forward and backward) hash chains as shown in Fig. \ref{fig:SessionKey}.
For the forward chain, both SM$_{i}$ and $\mathcal{P}_{i,j}$ compute $F_1=H_{1}(K_{i,j}, TS, 1)$ where $TS$ is the timestamp.
Then, all the elements of the forward hash chain are computed using $F_{a}=H_{1}(F_{a-1})$ for $2 \leq a \leq T$, where $T$ is the number of short-term keys that need to be stored.
%For increased security, each key is only used for one message
Similarly, the backward hash chain is computed by first computing $B_T=H_{1}(K_{i,j}, TS, 2)$, then the elements of the backward hash chain are computed as $B_{b}=H_{1}(B_{b+1})$ for $1 \leq b \leq T-1$.
Finally, the short-term key is computed by XORing an element from the forward chain with the corresponding element from the backward chain and hashing the result as shown in the figure.
Each short-term key should be used for a short time and after using all the $T$ keys, the meters can compute a new set of keys using an updated TS.
In this way, the smart meters do not need to compute and store a large number of keys in each round.
Short-term keys are computed daily and each key should be used to compute a set of masks.
%one mask that is used for reporting one reading.
After using the long-term key for a certain time, SM$_{i}$ and $\mathcal{P}_{i,j}$ should establish a new long-term seed key, and derive a new set of short-term keys.

%\begin{figure*}[!t]
%\centering \includegraphics[scale=0.38]{Recover_cropped} \caption{Example of exchanged, stored messages in our scheme and bottom-up aggregation.\label{fig:Recover} }
%\vspace{-4mm}
%\end{figure*}

%\subsubsection{Keys synchronization}
%%In our aggregation scheme,
%SM$_{i}$ and $\mathcal{P}_{i,j}$ should use the same key to compute the same mask so that masks cancel each other after aggregation.
%%If SM$_{i}$ and $\mathcal{P}_{i,j}$ lose mask synchronization such that the mask added by SM$_{i}$ is different from the mask added by $\mathcal{P}_{i,j}$, the correct aggregated reading cannot be recovered. \hl{Pollution could come to the reviewer mind later}.
%Losing key synchronization is unlikely in our scheme because each key should be used for a predefined number of readings, however, to enable the utility to identify a lack of key synchronization if it happens, a known value $\epsilon$ should be sent frequently by every SM as a reading, e.g., after sending a certain number of readings.
%This value undergoes the process of in-network aggregation as usual.
%Then, the utility should receive an aggregated reading of $n\epsilon$, where $n$ is the total number of
%SMs in the network.
%If the aggregated reading is not $n\epsilon$, the utility should broadcast an error message to all SMs so that they can re-synchronize the keys by skipping the current set of short keys and move to the next set.

\begin{figure*}[!t]
	\centering \includegraphics[width=0.85\linewidth]{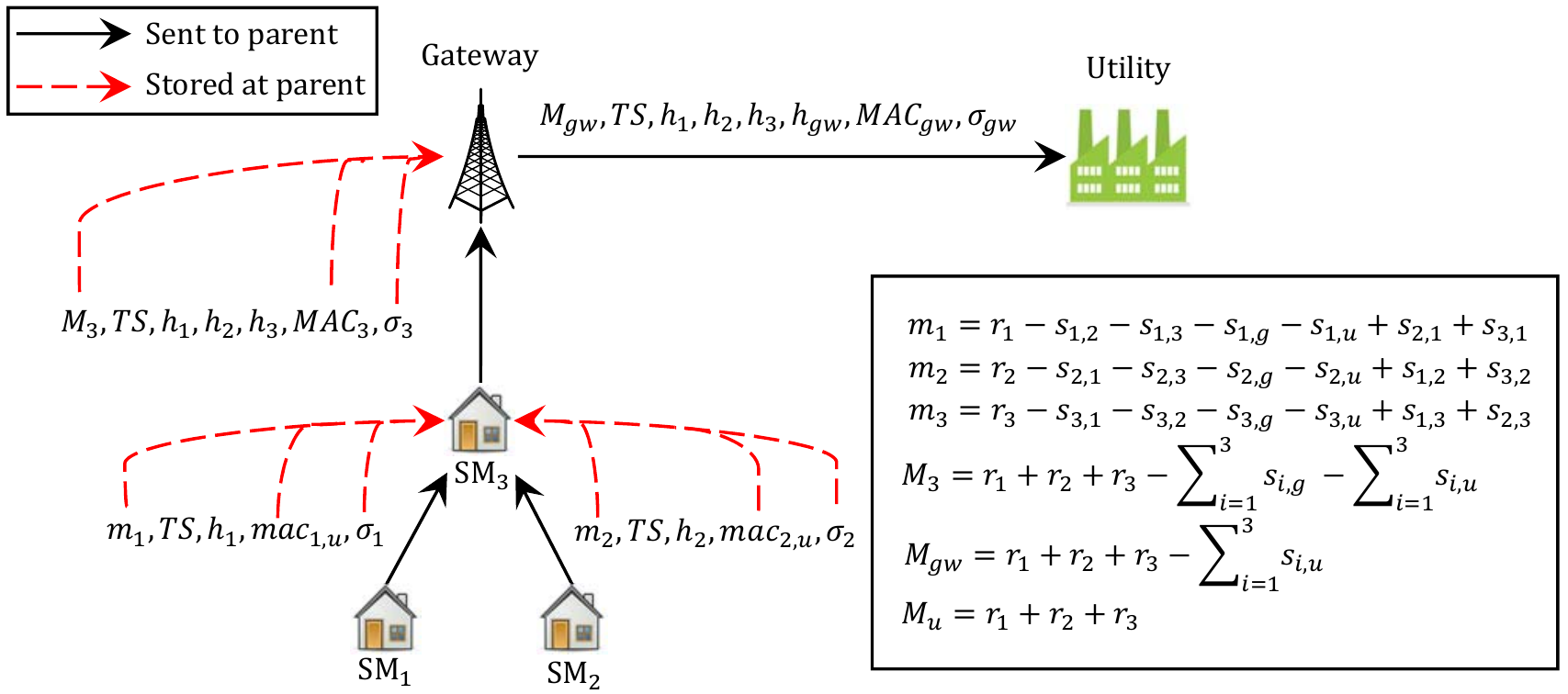}
	\caption{Example for exchanged messages, stored data, and bottom-up data aggregation.\label{fig:Recover} }
	\vspace{-4mm}
\end{figure*}

\section{The Proposed Scheme}\label{sec:Proposed-Scheme}
In this section we give the details of EPIC starting by system setup.
Then, we show how SMs report and aggregate power consumption readings.
Finally, we illustrate how the utility can recover the aggregated reading, verify its integrity, users' authenticity, and generate electricity
bills for each user.

%
%Fig. \ref{fig:Recover} shows an example of a network that has only three meters (SM$_{1}$, SM$_{2}$ and SM$_{3}$), the gateway, and a utility.
%To simplify our description, we assume that the proxies of each meter are all the other meters, gateway, and the utility.
%As shown in the figure, leaf meters, SM$_{1}$ and SM$_{2}$, compute the masked readings,$m_1$ and $m_2$, and generate a report to be sent to the parent meter, SM$_3$.
%SM$_3$ uses the received homomorphic hashes and mac values to check the received messages integrity, then, it generate its masked reading $m_{3}$, compute the aggregated reading $M_{3}$ , and finally submit a report to the gateway.
%As shown in the figure, during the aggregation process by SM$_{3}$, the masks $s_{1,2},s_{1,3},s_{2,3},s_{2,1},s_{3,1},s_{3,2}$ were removed.
%The remaining masks are removed incrementally such that when the aggregated reading reach the utility, it contains only masks that can be removed by the utility.
%Non-leaf nodes, SM$_3$ and the gateway, store the received masked readings, mac values and signature.
%If the utility detects a data modification attack, it requests all the data stored by non-leaf nodes to identify the attacker.

\subsection{System Setup}

\label{sec:System setup} \vspace{0mm}

An offline trusted authority (TA) should bootstrap the system as follows.
First, the TA generates the bilinear mapping parameters $\left(q,\mathbb{G}_{1},\mathbb{G}_{2},P,\hat{e}\right)$.
%In addition, the TA chooses symmetric key encryption and decryption schemes $E_K(.)$ and $D_K(.)$, where $K$ is the key.
It also chooses three different hash functions.
$H_{1}$ is a regular  hash function such as SHA-1, $H_{2}:\left\{0,1\right\}^{*}\rightarrow\mathbb{G}_{1}$, and $\mathcal{H}$ is a homomorphic hash function with the generators $\left\{ P_{1},P_{2},...,P_{d}\right\}\in\mathbb{G}$.
Furthermore, a keyed hash function $mac_{i,u}\leftarrow HMAC_{K_{i,u}}\left(h\right)$ is selected, where $mac_{i,u}$ is the HMAC on $h$ using the symmetric key $K_{i,u}$.
Then, the TA publishes the system public parameters as
$pubs=\{q, P, P_{1}, P_{2}, ..., P_{d}, \mathcal{H},$ $H_{1}, H_{2}, HMAC_{K}, \hat{e}\}$.
%$pubs=\{q, P, P_{1}, P_{2}, ..., P_{d}, \mathcal{H},$ $H_{1}, H_{2}, HMAC_{K}, \hat{e}, E_K(.), D_K(.)\}$.

In addition, every SM$_{i}$ chooses a secret key $x_{i}$ $\in$ $\mathbb{Z}_{q}^{*}$ and computes the corresponding public key $Y_{i}=x_{i}P$.
It should also obtain a certificate for the public key from a certificate authority.
Finally, each SM$_{i}$ should select several proxies, assuming that SM$_{i}$ selects $\alpha_{i}$ proxies and be selected by $\beta_{i}$ meters to act as a proxy for them, i.e. the total number of proxies for SM$_i$ is $\lambda_i$$=$$\alpha_i$$+$$\beta_i$.
Each SM and its proxies should establish the long-term seed key, derive the short-term keys, and compute the shared masks  as explained in Section \ref{sec:aggregation}.

\subsection{Leaf Meters: Report Generation} \label{subsec:Leaf_node}
Each leaf meter SM$_c$ (like SM$_1$ and SM$_2$ in Fig. \ref{fig:Recover}) generates a power consumption report by executing the following steps.

\begin{enumerate}
	\item Masks its reading $r_c$ to obtain a masked reading $m_c$.
	\begin{equation} \label{eq:leaf_masked_reading}
	m_c=r_{c}-\sum_{j=1}^{\alpha_{c}}{s_{c,j}}+\sum_{j=1}^{\beta_{c}}{s_{j,c}}
	\end{equation}
	\item Hashes its masked reading $m_c$ using homomorphic hash function $\mathcal{H}(\ )$ to get $h_c$.
	\begin{equation} \label{eq:leaf_hash}
	h_c=\mathcal{H}\left(m_c\right)\equiv\mathcal{H}\left(r_c\right)-\sum_{j=1}^{\alpha_{c}}{\mathcal{H}\left(s_{c,j}\right)}+\sum_{j=1}^{\beta_{c}}{\mathcal{H}\left(s_{j,c}\right)}
	\end{equation}
	
	\item Computes HMAC on $h_c$ using the shared key with the utility as
	$mac_{c,u}=\text{HMAC}_{K_{c,u}}\left(h_{c}\right)$.
	
	\item Generates a signature
	$\sigma_{c}=x_{c}H_{2}\left(m_{c}, mac_{c,u}, TS\right)$.
\end{enumerate}
Finally, SM$_c$ transmits to its parent SM$_{i}$ the following tuple
\begin{equation}\label{eq:leaf_report}
%	E_{K_{c,i}}\left(m_c||TS||h_c\right), mac_{c,u}, \sigma_{c}
m_c, TS, h_c, mac_{c,u}, \sigma_c
\end{equation}

\subsection{Non-leaf Nodes: Data Verification and Report Generation}
\label{subsec:Report-Generation-and}

The operations done by non-leaf meters SM$_i$ (like SM$_3$ in Fig. \ref{fig:Recover}) and the gateway  can be divided into two phases.
In the first phase, SM$_{i}$ receives $n_{i}$ messages from its children and verifies the authenticity and integrity of the received messages.
In the second phase, SM$_{i}$ create a new message to be transmitted to the next parent.
%aggregates the received masked readings with its reading and reports the aggregation to its parent.
These two phases should be executed at each non-leaf node until the aggregated masked reading reaches the utility.
The details of the two phases are as follows.

\textit{Phase 1.}
SM$_{i}$ receives $n_{i}$ messages from each child meter SM$_c$ $(1\leqslant c\leqslant n_{i})$.
If the child is a leaf-node, its message has this format $\left(m_{c}, TS, h_c, mac_{c,u}, \sigma_{c}\right)$ while if the child is a non-leaf node, the message has the following format %$\left(E_{K_{c,i}}\left(M_c||TS||h_{1},h_{2},\dots,h_{\ell_c}\right),MAC_{c},\sigma_{c}\right)$
$\left(M_c, TS, h_{1}, h_{2}, \dots, h_{\ell_c}, MAC_{c}, \sigma_c\right)$
where $M_c$ and $MAC_c$ are aggregated masked reading and aggregated MAC computed by the non-leaf chiled SM$_c$ as defined in Table \ref{tab:Symbols-and-Abbreviations}.
Also, the message contains the hashes of the masked readings of the sub-tree nodes of child SM$_c$.
%Note that, if the child is a leaf-node, its message has this format $\left(m_{c}||TS||h_c||mac_{c,u}||\sigma_{c}\right)$.
%$\left(E_{K_{c,i}}\left(m_{c}||TS||h_{\ell_c}\right),mac_{c,u},\sigma_{c}\right)$.
SM$_{i}$ should perform the following verifications.
% decrypts $E_{K_{c,i}}\left(M_c||TS||h_{1},h_{2},\dots,h_{\ell_c}\right)$ and
\begin{enumerate}
	
	%\item Verify that the packets are not stale by checking the timestamps $TS$ to thwart replay attacks.
	
	\item Perform a batch verification for the received signatures to make sure that the packets are sent from the intended meters as follows.
	
	\begin{equation}
	\hat{e}\left(\sum_{c=1}^{n_{i}}\sigma_{c},P\right)\stackrel{?}{=}\prod_{c=1}^{n_{i}}\hat{e}\big(H_{2}\left(M_{c}, MAC_{c}, TS\right),Y_{c}\big)
	\label{eq:sign}
	\end{equation}
	
	%\begin{equation}
	%\begin{aligned}\hat{e}\left(\sum_{c=1}^{n_{i}}\sigma_{c},P\right) & \stackrel{?}{=}\prod_{c=1}^{n_{i}}\hat{e}\left(x_{c}H_{2}\left(M_{c}||MAC_{c}||TS\right),P\right)\\
	% & =\prod_{c=1}^{n_{i}}\hat{e}\big(H_{2}\left(M_{c}||MAC_{c}||TS\right),x_{c}P\big)\\
	% & =\prod_{c=1}^{n_{i}}\hat{e}\left(H_{2}\left(M_{c}||MAC_{c}||TS\right),Y_{c}\right)
	%\end{aligned}
	%\label{eq:sign}
	%\end{equation}
	
	\item Perform a batch verification for all the received hashes by checking
	\begin{equation}
	\mathcal{H}\left(\sum_{c=1}^{n_{i}}M_{c}\right)\stackrel{?}{=}\sum_{c=1}^{n_{i}}\sum_{j=1}^{\ell_{c}}h_{j}
	\label{eq:verify}
	\end{equation}
	If this verification passes, SM$_i$ moves to the next step, otherwise,
	%If the verification fails, then
	data modification attack is detected and SM$_i$ can identify the attacker by applying divide-and-conquer verification recursively until the attacker is identified.

	%using the procedure explained later in subsection \ref{sec:Attacks against data integrity}.
	
	\item Store the latest received tuple $\left(M_{c},MAC_{c},\sigma_{c}\right)$ from every child to help the utility to identify the attacker in case that the utility detects data modification attack, as will be explained in the next subsection.
\end{enumerate}

\textit{Phase 2.} In this phase, SM$_{i}$ should execute the following steps before sending a reading packet to its parent.
\begin{enumerate}
	\item Masks its fine-grained reading $r_{i}$ to obtain its own masked reading $m_i$
	\begin{equation}\label{eq:non_leaf_masked_reading}
	m_{i}=r_{i}-\sum_{j=1}^{\alpha_{i}}{s_{i,j}}+\sum_{j=1}^{\beta_{i}}{s_{j,i}}
	\end{equation}
	
	\item Aggregates its masked reading $m_i$ with the masked readings received from its children meters to generate an aggregated masked readings $M_i$ as
	\begin{equation}\label{eq:non_leaf_aggregation}
	M_{i}=m_{i}+\sum_{c=1}^{n_{i}}M_{c}
	\end{equation}

	\item Hashes its masked reading $m_i$ using homomorphic hash function to get $h_{i}$ as
	\begin{equation}\label{eq:non_leaf_hash}
	h_{i}=\mathcal{H}\left(m_{i}\right)\equiv\mathcal{H}\left(r_i\right)-\sum_{j=1}^{\alpha_{i}}{\mathcal{H}\left(s_{i,j}\right)}+\sum_{j=1}^{\beta_{i}}{\mathcal{H}\left(s_{j,i}\right)}
	\end{equation}
	
	\item Computes HMAC on $h_i$ with the shared key with the utility as
	$mac_{i,u}=\text{HMAC}_{K_{i,u}}\left(h_{i}\right)$.

	\item Aggregates its MAC with the received aggregated MACs using XOR operations to obtain $MAC_{i}=mac_{i,u}\oplus\left(\bigoplus_{c=1}^{n_{i}}MAC{}_{c}\right)$.
	\item Generates a signature
	$\sigma_{i}=x_{i}H_{2}\left(M_{i}, MAC_{i}, TS\right)$.
	
\end{enumerate}
Finally SM$_{i}$ sends to its parent
SM$_{pi}$
the following tuple
\begin{equation}\label{eq:non_leaf_report}
M_{i}, TS, h_{1}, h_{2}, \dots, h_{\ell_{i}}, MAC_i, \sigma_i
%E_{K_{i,pi}}\left(M_{i}||TS||h_{1},h_{2},\dots,h_{\ell_{i}}\right),MAC_{i},\sigma_{i}
\end{equation}

This process of verification and aggregation proceeds in a bottom-up manner to the utility.

\begin{table*}[tp]
	\caption{Addition of $w$ masked reports of each meter to obtain the total power consumption during billing period.
		\label{tab:Billing-and-dynamic}}
	\vspace{-2mm}
	\center
	\resizebox{0.99\textwidth}{0.12\textwidth}{ %
		\begin{tabular}{cccccc}
			\toprule
			& \textbf{\textit{$t_{1}$}} & \textit{...}  & \textbf{\textit{$t_{w-1}$}}\textbf{ } & \textbf{\textit{$t_{w}$}}\textbf{ } & \multicolumn{1}{c}{\textbf{\textit{Billing Period Consumption}}}\tabularnewline
			\midrule
			\textbf{SM$_{1}$}\textbf{ } & $r_{1}^{(1)}+\sum_{j=1}^{\alpha_{1}}s_{1,j}^{(1)}-\sum_{j=1}^{\beta_{1}}s_{j,1}^{(1)}$ & \textit{...}  & $r_{1}^{(w-1)}+\sum_{j=1}^{\alpha_{1}}s_{1,j}^{(w-1)}-\sum_{j=1}^{\beta_{1}}s_{j,1}^{(w-1)}$ & $r_{1}^{(w)}+s_{1,u}^{(b)}-\sum_{k=1}^{w-1}(\sum_{j=1}^{\alpha_{1}}s_{1,j}^{(k)}-\sum_{j=1}^{\beta_{1}}s_{j,1}^{(k)})$ & \multicolumn{1}{c}{$s_{1,u}^{(b)}+\sum_{k=1}^{w}r_{1}^{(k)}$}\tabularnewline
			\vdots{} & \vdots{} &  & \vdots{} & \vdots{} & \vdots{}\tabularnewline
			\textbf{SM$_{n}$}\textbf{ } & $r_{n}^{(1)}+\sum_{j=1}^{\alpha_{n}}s_{n,j}^{(1)}-\sum_{j=1}^{\beta_{n}}s_{j,n}^{(1)}$ & \textit{...}  & $r_{n}^{(w-1)}+\sum_{j=1}^{\alpha_{n}}s_{n,j}^{(w-1)}-\sum_{j=1}^{\beta_{n}}s_{j,n}^{(w-1)}$ & $r_{n}^{(w)}+s_{n,u}^{(b)}-\sum_{k=1}^{w-1}(\sum_{j=1}^{\alpha_{n}}s_{n,j}^{(k)}-\sum_{j=1}^{\beta_{n}}s_{j,n}^{(k)})$ & $s_{n,u}^{(b)}+\sum_{k=1}^{w}r_{n}^{(k)}$\tabularnewline
			&  &  &  &  & \tabularnewline
			$gw$ & $\sum_{j=1}^{\alpha_{gw}}s_{gw,j}^{(1)}-\sum_{j=1}^{\beta_{gw}}s_{j,gw}^{(1)}$ & \textit{...}  & $\sum_{j=1}^{\alpha_{gw}}s_{gw,j}^{(w-1)}-\sum_{j=1}^{\beta_{gw}}s_{j,gw}^{(w-1)}$ & $-\sum_{k=1}^{w-1}(\sum_{j=1}^{\alpha_{gw}}s_{gw,j}^{(k)}-\sum_{j=1}^{\beta_{gw}}s_{j,gw}^{(k)})$ & \tabularnewline
			&  &  &  &  & \tabularnewline
			$u$ & $\sum_{j=1}^{\alpha_{u}}s_{u,j}^{(1)}-\sum_{j=1}^{\beta_{u}}s_{j,u}^{(1)}$ & \textit{...}  & $\sum_{j=1}^{\alpha_{u}}s_{u,j}^{(w-1)}-\sum_{j=1}^{\beta_{u}}s_{j,u}^{(w-1)}$ & $-\sum_{k=1}^{w-1}(\sum_{j=1}^{\alpha_{u}}s_{u,j}^{(k)}-\sum_{j=1}^{\beta_{u}}s_{j,u}^{(k)})$ & \multicolumn{1}{c}{}\tabularnewline
			\midrule
			Total & $\sum_{i=1}^{n}r_{i}^{(1)}$ & \textit{...}  & $\sum_{i=1}^{n}r_{i}^{(w-1)}$ & $\sum_{i=1}^{n}s_{i,u}^{(b)}+\sum_{i=1}^{n}r_{i}^{(w)}$ & \tabularnewline
			\bottomrule
		\end{tabular}
		
	}\vspace{-3mm}
\end{table*}

\subsection{Utility: Aggregated Reading Recovery, Data Integrity Verification and Billing}\label{sub:utility_operations}
\textit{Data recovery and verification}.
The utility receives %$(E_{K_{gw,u}}\left(M_{gw}||TS||h_{1},h_{2},\dots,h_{\ell_{gw}}\right),MAC_{gw},\sigma_{gw})$
$\left(M_{gw}, TS, h_1, h_2, \dots, h_{\ell_{gw}}, MAC_{gw}, \sigma_{gw}\right)$
from the gateway.
The utility first verifies $\sigma_{gw}$, homomorphic hashes, and TS, as described in \emph{phase 1} of subsection \ref{subsec:Report-Generation-and}.
Then, it verifies the aggregated $MAC_{gw}$ as follows.
\begin{enumerate}
	\item Calculates all the individual MACs from the received hashes
	$\left\{mac_{u,j}'=HMAC_{K_{u,j}}\left(h_j\right),\ \forall j\in\left\{ 1..\ell_{gw}\right\}\right\} $,
	
	\item Calculates the aggregated MAC $MAC_{u}'$=$\bigoplus_{j=1}^{\ell_{g}}mac_{u,j}'$.
	
	\item Compares the calculated MAC with received MAC.
	\begin{equation} \label{eq:mac_ver}
	MAC_{u}'\stackrel{?}{=}MAC_{gw}
	\end{equation}
\end{enumerate}
If the verification passes, the utility can recover the aggregated reading of all SMs by removing its masks from $M_{gw}$ as follows.
	\begin{equation} \label{eq:agg_read_final}
M_{gw}+\sum_{j=1}^{\beta_{u}}{s_{j,i}}=\sum_{i=1}^{n}r_{i}
\end{equation}

where $n$ is the total number of meters in the AMI network.
%If any message in the transition to the utility has been modified

\textit{Attacker identification}.
If a smart meter SM$_i$ modifies both the aggregated masked reading and a homomorphic hash of any child in its subtree, i.e., transmits $M_c'$ and $h_c'$ instead of $M_c$ and $h_c$ to bypass its parent verification done in Equation \ref{eq:verify}, then the utility verification done in Equation \ref{eq:mac_ver} fails because $MAC_{c}$ was computed by SM$_c$ on $h_c$ not $h_c'$.
%In this case the utility can identify the attacker using the procedure explained in subsection \ref{sec:Attacks against data integrity}.
In this case, data modification attack is detected and the utility suspects all non-leaf SMs since any non-leaf SM can launch this attack.
Therefore, the utility runs the following verifications in a bottom-up manner, i.e., starting from the first non-leaf nodes up to the last non-leaf node which is the gateway, until the attacker is identified.

In order to identify the attacker, the utility should retrieve SM$_i$ children reports  $\left(M_{c},MAC_{c},\sigma_{c}\right)$ $1 \leq c \leq n_i$ from SM$_{i}$ and
$\left(M_{i},MAC_{i},\sigma_{i}\right)$ from SM$_{pi}$, which is the parent of SM$_{i}$.
Then, the utility check if
\begin{equation}
\hat{e}\left(\sum_{c=1}^{n_{i}}\sigma_{c},P\right)\stackrel{?}{=}\prod_{c=1}^{n_{i}}\hat{e}\left(H_{2}\left(M_{c}, MAC_{c}, TS\right),Y_{c}\right)
\label{eq:inbound_check}
\end{equation}
If SM$_i$ can provide valid signatures from its children, it can pass the verification done in equation \ref{eq:inbound_check}, otherwise, SM$_i$ is identified as an attacker.
The attacking SM can pass this verification iff he sends the correct $M_c$ not $M_c'$, however, it will be identified by the next verification process.

If the verification in equation \ref{eq:inbound_check} passes for all non-leaf nodes, then the utility should to check the correctness of the messages sent by each meter SM$_i$.
First, the utility extracts its individual masked reading from the aggregated masked reading $M_i$ using
\begin{equation*}
m_i'=M_i-\sum_{i=c}^{n_{i}}M_{c}
\end{equation*}
Then, it re-calculates $mac_{i,u}$ from the verified masked readings as
\begin{equation*}
mac_{i,u}'=\text{HMAC}_{K_{i,u}}\left(\mathcal{H}\left(m_i'\right)\right)
\end{equation*}
After that, the utility re-calculates $mac_{i,u}$ from the verified aggregated MACs as
\begin{equation*}
mac_{i,u}''=MAC_{i}\oplus\left(\bigoplus_{c=1}^{n_{i}}MAC{}_{c}\right)
\end{equation*}
Finally the utility checks if
\begin{equation}
mac_{i,u}'\stackrel{?}{=}mac_{i,u}''
\label{eq:ougoing_check}
\end{equation}
If the verification fails, SM$_{i}$ is identified as an attacker.
This process continues in a bottom-up manner until the attacker is identified.
The attacker cannot pass this check because he cannot compute a valid mac value for the modified packet.
This is because the computation of a valid mac value requires the knowledge of the shared key between the victim meter, SM$_c$, and the utility.
Therefore, EPIC can ensure E2E data integrity and authenticity without accessing the fine-grained readings to preserve consumers' privacy.

\textit{Dynamic-pricing-based billing\label{sec:Billing and dynamic pricing}}.
For dynamic pricing, the utility can divide the day into periods with different electricity prices.
%Referring to the California duck curve shown in Fig. \ref{fig:CaliforniaDuckCurve}, every day can be divided into three periods. The low-price period (08:00 a.m. - 06:00 p.m.), the normal-price period (12:00 a.m. - 08:00 a.m.), and the high-price period (06:00 p.m. - 12:00 a.m.).
Assuming that the meters should report $w$ power consumption readings during each billing period,
Table \ref{tab:Billing-and-dynamic} gives the $w$ masked readings generated by $n$ meters,
where each column represents the masked readings sent in one time slot,
while each row represents all the masked readings sent by each meter during the billing period (i.e., $w$ readings).
As explained earlier, the reading $r_i$ of SM$_i$ at a time slot $t_{x}$ is masked using the mask $\sum_{j=1}^{\alpha_{i}}s_{i,j}^{(t_x)}-\sum_{j=1}^{\beta_{i}}s_{j,i}^{(t_x)}$ to produce the masked reading $m_i=r_i+\sum_{j=1}^{\alpha_{i}}s_{i,j}^{(t_x)}-\sum_{j=1}^{\beta_{i}}s_{j,i}^{(t_x)}$.
The masks should be computed in such a way that the summation of all the masks used during billing period is zero.
This can be done as follows.
At the end of each billing period (report at $t_{w}$), the mask SM$_i$ should use is equal to the negative summation of all the previous $w-1$ masks plus a billing mask, $s_{i,u}^{(b)}$, shared between the meter and the utility, i.e., the mask that should be used in the last reading of a billing period is
\[
s_{i,u}^{(b)}-\sum_{k=1}^{w-1}\left(\sum_{j=1}^{\alpha_{i}}s_{i,j}^{(t_k)}-\sum_{j=1}^{\beta_{i}}s_{j,i}^{(t_k)}\right)
\]
so that the summation of all masked readings of SM$_i$ gives the total power consumed by SM$_{i}$ plus the billing mask, i.e.,
\[
\sum_{k=1}^{w}m_{i}^{(k)}=\sum_{k=1}^{w}r_{i}^{(k)}+s_{i,u}^{(b)}
\]

The utility should compute $\sum_{k=1}^{w}r_{i}^{(k)}$ to bill SM$_i$.
It can use the homomorphic hash property
$\mathcal{H}(m_{1}+m_{2})=\mathcal{H}(m_{1})+\mathcal{H}(m_{2})$ to compute $\sum_{k=1}^{w}r_{i}^{(k)}$ as follows.
First, the utility should add all the $w$ homomorphic hashes sent by SM$_i$ in the billing period to obtain
%which is equivalent to a the homomorphic hash of the total power conumed by SM$_i$ during the billing period
\\
\[
\sum_{t_x=1}^{w}h_{i}^{(t_x)}=\mathcal{H}\left(\sum_{k=1}^{w}r_{i}^{(k)}\right)+\mathcal{H}\left(s_{i,u}^{(b)}\right)
\]
%\[
%$\sum_{t=1}^{w}h_{i}^{(t_x)}=\mathcal{H}\left(\sum_{k=1}^{w}r_{i}\right)=\mathcal{H}\left(R_{i}\right)\
%\]
It is clear that only the utility can remove $\mathcal{H}\left(s_{i,u}^{(b)}\right)$ and hence only the utility can obtain $\mathcal{H}\left(\sum_{k=1}^{w}r_{i}^{(k)}\right)$.
Since the range of the readings is small, the utility can build a look-up table and obtain the total power consumption of SM$_i$, $\sum_{k=1}^{w}r_{i}^{(k)}$, from $\mathcal{H}\left(\sum_{k=1}^{w}r_{i}^{(k)}\right)$.
It should be noted that, it is easy to obtain $\sum_{k=1}^{w}r_{i}^{(k)}$ from $\mathcal{H}\left(\sum_{k=1}^{w}r_{i}^{(k)}\right)$ since all the masks are canceled and the total consumption of a billing period is not a large number, but it is extremely hard to obtain $m_i$ from $h_i$ since the masks can make $m_i$ a large number. % (128 bit).
Knowing the power consumption of SM$_i$ during the billing period does not degrade consumers' privacy because the time period is long enough to prevent sensitive data leakage \cite{PP11}.

\section{Security and Privacy Analysis} \label{sec:Security-and-privacy}

In this section, we analyze the security properties of EPIC.
In specific, our analysis focuses on how our scheme can collect power consumption readings with privacy preservation, collusion resistance, authenticity, and E2E integrity verification of the aggregated reading and identification of attackers.

%In this section, we explain the possible security attacks that can target our model. We will also explain how our scheme can thwart these attacks.

\subsection{Privacy Analysis}

\subsubsection{Singular Attacks}
%We consider in this part
An adversary, $\mathcal{A}$, can eavesdrop on all the communications of all the network nodes and can obtain the individual masked readings of a leaf SM.
However, based on Equation \ref{eq:leaf_masked_reading}, $\mathcal{A}$ must know all the $\lambda_c$ masks, $\lambda_c=\alpha_c+\beta_c$, shared between the leaf meter, SM$_c$, and its proxies to be able to extract the meter's fine-grained reading.
Since no entity can compute the correct masks except SM$_c$ and its proxies, as explained in subsection \ref{sec:Key agreemen}, $\mathcal{A}$ cannot obtain the meters' fine-grained readings.

In the following, we present a formal security proof to show that the masking technique used in EPIC is semantic secure against chosen-plaintext attacks even if only one mask value is used to mask the fine-grained reading.

%Consider the proposed masking technique in EPIC and chosen-plaintext attacks on EPIC according to the following game.

\begin{thm}
	The masking scheme is semantically secure against chosen-plaintext attacks under the pseudorandom function (PRF) assumption for HMAC.
\end{thm}

\begin{IEEEproof}
	The theorem proof as a game is constructed as follows.
	\begin{itemize}
		%\item \textbf{Initiallization}. The challenger $\mathcal{C}$ is initiated with a set of short-term keys where each key is used only once to generate a one-time secret mask using the HMAC as pseudorandom function (PRF) \cite{HMAC_Security}.
		
		\item \textit{Initialization}: The challenger $\mathcal{C}$ is initiated with a set of one-time secret masks generated as explained in \autoref{sub:Masking} using the HMAC function which is used as a PRF \cite{HMAC_Security}.	
		
		\item \textit{Challenge}: The adversary $\mathcal{A}$ outputs two fine-grained readings $r_0$ and $r_1$ to $\mathcal{C}$.
		$\mathcal{C}$ chooses a random bit $b\in\{0,1\}$ and responds with a ciphertext $m_b=Enc(r_b)=r_b+s$, where $s$ is the one-time secret mask and $s\gg r_b$.
		
		\item \textit{Guess}: The adversary $\mathcal{A}$ responds with $b'\in\{0,1\}$ as a guess for $b$. The advantage of the adversary against the masking  in the above game can be defined as
		\begin{equation*}
		\text{Adv}_{\mathcal{A}}=\left|\text{Pr}[b'=b]-\frac{1}{2}\right|
		\end{equation*}
		Because $s\gg r_b$, $s$ is generated by a PRF and used only one time, the advantage $\text{Adv}_{\mathcal{A}}$  in this case becomes
		\begin{equation*}
		\text{Adv}_{\mathcal{A}}=\left|\frac{1}{2}-\frac{1}{2}\right|=0
		\end{equation*}
		Therefore, the masking scheme is semantically secure and no adversary can extract the fine-grained reading.
	\end{itemize}
	
\end{IEEEproof}

In addition, each mask value is used for only one reporting period to ensure that the masked readings look different even if the leaf meter reports the same fine-grained reading at different time slots.
Therefore, given two consecutive reports of a meter, $\mathcal{A}$ can not learn if the power consumption has changed or not.

% ===========================================================================================

\subsubsection{Collusion attacks}
Unlike the singular attacks launched by a single adversary, we consider in the following a stronger attack in which the adversary can collude with other nodes in the AMI networks.

In EPIC, the fine-grained reading of each meter is protected by $\lambda$ secret masks shared with $\lambda$ proxies.
Therefore, for an attacker to compute the fine-grained reading of a victim meter, \textit{the attacker must collude at least with all the victim's proxies}.
In particular, the attacker can try to recover the fine-grained reading of a victim meter from either its masked reading or its homomorphic hash.

To recover the fine-grained reading of a victim leaf meter from its masked reading, \textit{the attacker must collude with all the victim's proxies} to obtain all secret masks and use them to get the fine-grained reading from the masked reading given in Equation \ref{eq:leaf_masked_reading}.
If the victim is a non-leaf meter, based on Equation \ref{eq:non_leaf_aggregation}, \textit{the attacker must collude with both the victim's direct children and proxies}.

On the other hand, to recover the fine-grained reading of a victim meter from its homomorphic hash, based on Equation \ref{eq:leaf_hash} or \ref{eq:non_leaf_hash}, \textit{the attacker must collude with all the victim's proxies}
to remove the hashes of the secret masks and obtain the $\mathcal{H}(r_i)$ from $\mathcal{H}(m_i)$.
Since the value of $r_i$ is a small number, the attacker can build a look-up table and recover $r_i$ from $\mathcal{H}(r_i)$.

\begin{figure}
	\center
    \setlength{\abovecaptionskip}{-0.2cm} % 0.5cm as an example
	\setlength{\belowcaptionskip}{-1cm} % 0.5cm as an example
	\includegraphics[width=0.85\linewidth]{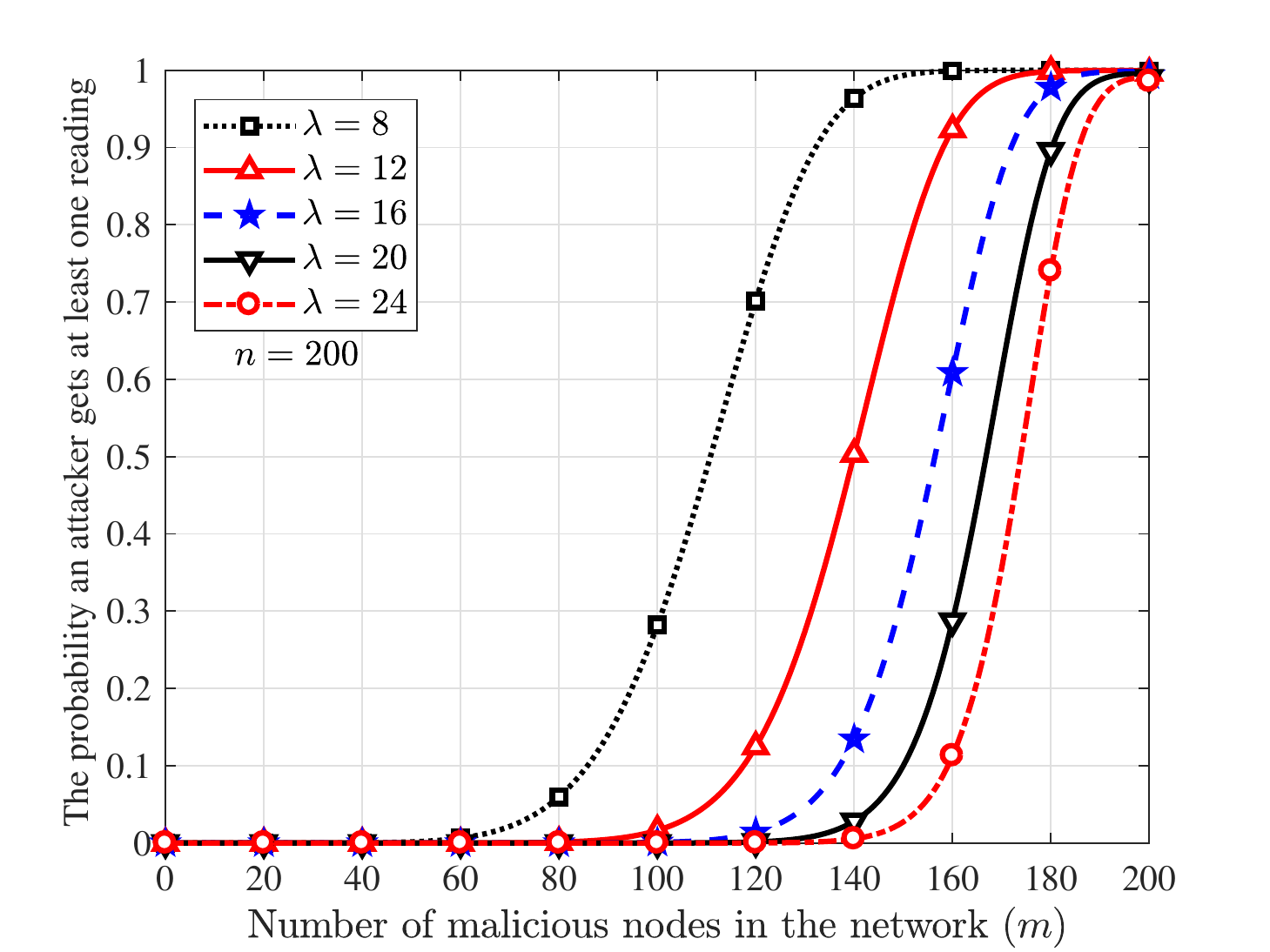}
	%\vspace{-1mm}
	\caption{The probability an attacker gets at least one reading for $n=200$.}
	\label{fig:collusion_analysis_attacker_side}
	\vspace{-5mm}
\end{figure}

In all the previous attack scenarios, \textit{the protection level against collusion attack is determined by the number of selected proxies} $\lambda$.
Therefore, in the following, we model an attack and investigate how a proper value for $\lambda$ can ensure a satisfactory protection level against collusion attack.

Consider that each SM selects $\lambda$ proxies, the network has $n$ nodes, including SMs, the gateway, and the utility, and the network has $m$ malicious nodes that collude with the attacker.
%Then, the probability that the fist selected proxy is malicious is $\frac{m}{n-1}$, and the probability that the second selected proxy is malicious is $\frac{m-1}{n-2}$.
The probability that a SM selects all the $\lambda$ proxies from the $m$ malicious nodes follows the hypergeometric probability distribution and is given by\vspace{-2mm}

\begin{equation*}
\frac{\Comb{m}{\lambda}}{\Comb{(n+1)}{\lambda}}
\end{equation*}
%\hlgreen{Collusion attack fails to recover the reading of the SM if at least one proxy does not collude, and }
Then, the probability that a meter is secure against collusion attack is  \vspace{-2mm}

\begin{equation*}
1-\frac{\Comb{m}{\lambda}}{\Comb{(n+1)}{\lambda}}
\end{equation*}
Let $\mathbb{P}$ be the probability that the attacker can recover at least the readings of any SM in the $(n-m)$ benign meters.  $\mathbb{P}$ can be expressed as\vspace{-2mm}

\begin{equation*}\label{eq:probability}
\mathbb{P}=1-\prod_{i=1}^{n-m}\left(1-\frac{\Comb{m}{\lambda}}{\Comb{(n+1)}{\lambda}}\right)
\end{equation*}

%the ratio of malicious nodes in the network $\frac{m}{n}$

To assess how hard for the attackers to launch successful collusion attack in EPIC, Fig. \ref{fig:collusion_analysis_attacker_side} gives $\mathbb{P}$ versus $m$ at different cases of $\lambda$ for $n=200$ SMs.
As shown in the figure, if each SM selects $\lambda=8$ proxies and 60 SMs colludes with the attacker, the probability that the attacker can obtain at least one meter's readings is almost zero.
The attacker needs to collude with 80 SMs, 40\% of the SMs in the network, so that the probability he can get at least one meter's readings becomes 0.06.
If each SM increases the level of protection against collusion attacks by adding $4$ more proxies, i.e. increasing $\lambda$ from $8$ to $12$, the SMs are almost secure when the attacker colludes with 80 SMs.
In this case, the attacker needs to collude with 120 SMs, 60\% of the network, so that $\mathbb{P}$ becomes 0.12.
We can conclude that, the increase of the number of proxies ($\lambda$) can make collusion attack harder to succeed.

\begin{figure}
   \setlength{\abovecaptionskip}{-0.2cm} % 0.5cm as an example
	\setlength{\belowcaptionskip}{-1cm} % 0.5cm as an example
	\center
	\includegraphics[width=0.85\linewidth]{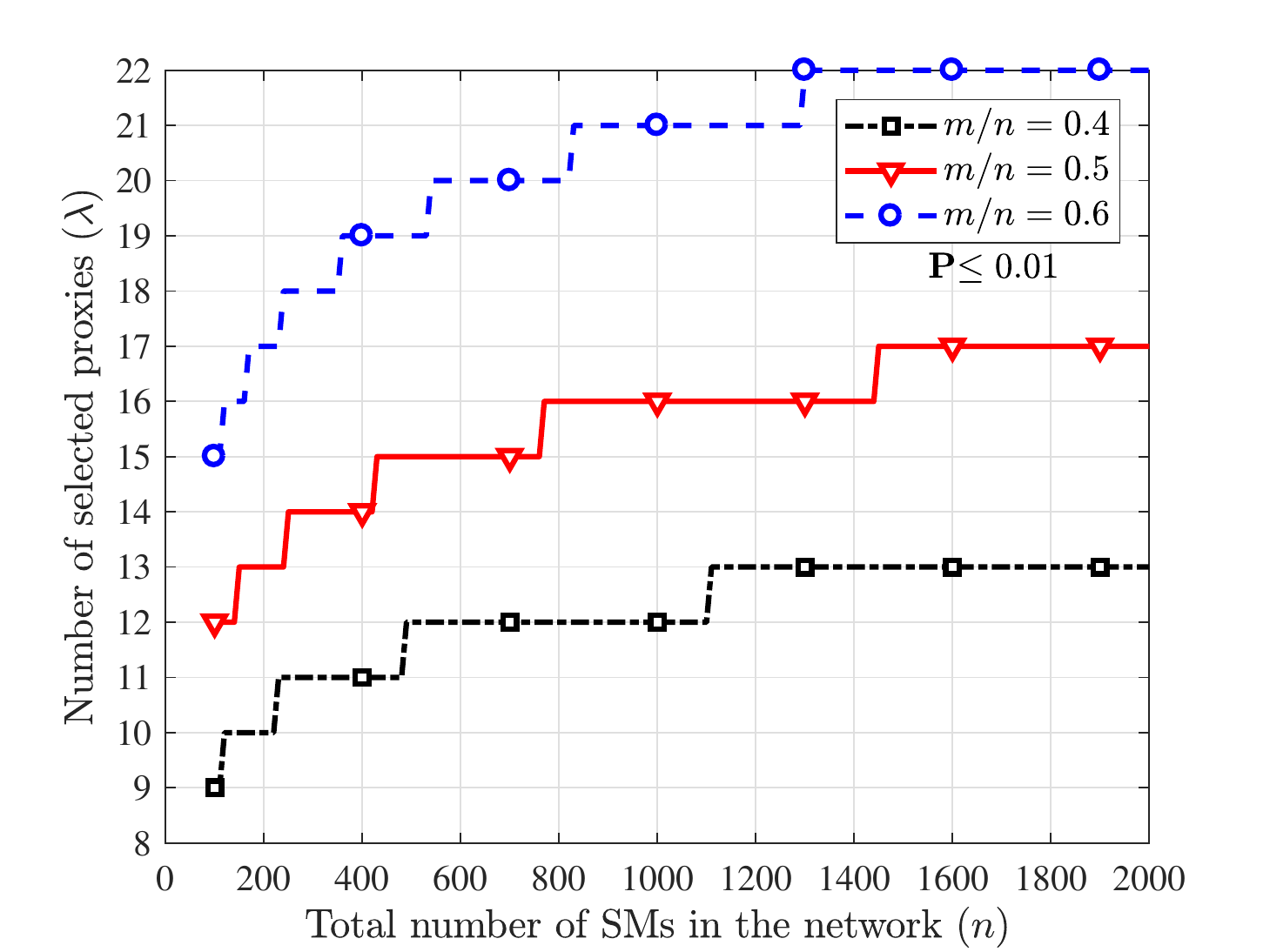}
	%\includegraphics[scale=0.51]{1.png}
%	\vspace{-2mm}
	\caption{Number of proxies vs. network size for $\mathbb{P}\leq0.01$.}
	\label{fig:collusion_analysis_selecting_proxies}
	\vspace{-5mm}
\end{figure}

%To show how SMs can be secured against collusion attacks in our scheme,
%To illustrate how many proxies should be selected by each SMs to be secured against collusion attacks in scalable and malicious networks, we plot in Fig. \ref{fig:collusion_analysis_selecting_proxies} $\lambda$ vs. $n$ such that $\mathbb{P}\leq0.01$.
To illustrate how many proxies should be selected by each SM to be secure against collusion attacks, Fig. \ref{fig:collusion_analysis_selecting_proxies} shows $\lambda$ vs. $n$ such that $\mathbb{P}\leq0.01$.
We define $m/n$ as the ratio of malicious nodes in the network.
%The figure concludes that, a small number of proxies, at most 1\% of the network,
A SM can select a proper number of proxies to be secure against collusion attack based on the network size and a risk assessment for the number of potential malicious meters in the network.
For example, for a network with 100 SMs and 40 SMs of them are malicious, a SM can be secured by selecting 9 proxies, whereas, if the network size increases to 2,000 SMs and 800 of them are malicious,
i.e., same $m/n$ ratio, the SM should increase the number of proxies from 9 to 12 to ensure that the probability of successful collusion attack is less than 0.01.
This indicates that although the number of SMs significantly increases from 100 to 2,000, a slight increase in the number of proxies is needed to secure the meters against collusion attacks.
%By increasing the network size from 100 SMs to 2000 SMs with half of the network are malicious nodes, a SM can be secured by increasing his proxies from 12 to 17, whereas
Moreover, in an extreme case in which $m/n=0.6$ and $n=2,000$, 22 proxies are needed to ensure that $P\leq0.01$.
We can conclude from this analysis that SMs can control the protection level against collusion attacks by selecting a proper number of proxies, and
the ratio of proxies to the network size ($\lambda/n$) is small to achieve a satisfactory protection against collusion attacks.
%, about 1\% of the network size for huge networks and can reach 10\% of the network size for smaller networks.

%To study the impact of the network size on tree data exposure in our scheme, we plot in Fig. \ref{fig:collusion_analysis_tree_size} $\mathbb{P}(q,\lambda,n)$ versus $q$ for different tree sizes at a fixed number of proxies $\lambda=25$ meters for each meter in each network.
%As shown in the figure, for a fixed number of proxies, increasing the network size, e.g. assuming scalable network, increases the probability of tree data exposure.
%However, a good selection of the number of proxies can ensure that the network scalability has a very small impact on tree data exposure.
%As shown in the figure, assuming $50/50$ probability of honest/malicious meters, selecting $\lambda=25$ meters can ensure that tree data exposure probability is $0$ if the network size is increased from $n=127$ meters to $n=2047$ meters.
%
%Therefore, our scheme has high resistance against collusion attacks with proper selection of the number of proxies per each meter despite the network size.

%\vspace{-0.8cm}
\subsection{Security Analysis}

\subsubsection{Data integrity} \label{sec:Attacks against data integrity}
If an external adversary $\mathcal{A}$ manipulates the transmitted messages between a child meter and its parent, the attack can be easily detected by the parent because it can verify the integrity of the received messages by verifying the received signature.
Forging a signature or modifying a valid signature is infeasible without knowing the private key of the child meter.
In addition, $\mathcal{A}$ may record valid packets exchanged between a meter and its parent (such as the packets given in (\ref{eq:leaf_report}) and (\ref{eq:non_leaf_report}) and replay them at later time to disrupt the reading collection scheme.
Since packets have timestamps, the stale packets can be easily identified and dropped.
If $\mathcal{A}$ tries to change the timestamp so that the packet looks fresh,
$\mathcal{A}$ needs to know the private key of the victim meter to compute a valid signature on the packet of the modified timestamp.

Comparing to external attackers, internal attacks can launch stronger attacks.
In particular, they may breach the data integrity by launching three different attacks:
(1) modification of a child's homomorphic hash only;
(2) modification of a child's masked reading only;
and (3) modification of both child's homomorphic hash and masked reading.
%To show how the attack can be detected and the attacker is identified, let SM$_c$ be the victim child, SM$_i$ be the malicious parent, and SM$_{pi}$ be the next parent which either detect the attack and identify the attacker, or help the utility in identifying the attacker.
Consider SM$_c$ be the victim child, SM$_i$ be the malicious parent, and SM$_{pi}$ be the parent of SM$_i$.
SM$_{pi}$ can either detect the attack of SM$_i$ or help the utility to detect the attack.
The first two attacks can be detected by SM$_{pi}$ because the batch verification process of the individual homomorphic hash values done by SM$_{pi}$ (given in Equation \ref{eq:verify}) fails.
%\begin{equation}
%\mathcal{H}\left(\sum_{i=1}^{n_{pi}}M_{i}\right)\stackrel{?}{=}\sum_{i=1}^{n_{pi}}\sum_{j=1}^{\ell_{i}}h_{j}
%\label{eq:verify2}
%\end{equation}
%If the equality does not hold, the attack is detected.
%In this case, SM$_{pi}$ can apply divide-and-conquer approach %\cite{d_and_c}
%recursively until it finds the malicious parent SM$_i$ for which
%\begin{equation*}
%\mathcal{H}\left(M_{i}\right)\neq\sum_{j=1}^{\ell_{i}}h_{j}
%%\label{eq:verify2}
%\end{equation*}
%and report the result with the proof of attacker identification to the utility.
For the third attack, modification of both $M_{c}$ and $h_{c}$, the utility can detect the attack from the aggregated MAC verification done in Equation \ref{eq:mac_ver}.
To identify the malicious SM$_i$, the utility should use the procedure explained in subsection \ref{sub:utility_operations}.
Therefore, EPIC can ensure E2E data integrity without accessing the fine-grained readings to preserve consumers' privacy.

\subsubsection{E2E users' Authenticity}
EPIC achieves hop-by-hop authentication in which each parent meter can authenticate the child meters because each packet is signed by the child meter.
Therefore, it is infeasible for $\mathcal{A}$ to impersonate meters by sending packets under their names, and thus parent meters accept only messages from authenticated children.
%In addition, the utility can make sure the readings are sent from the intended meters by verifying the aggregated MAC.
In addition, EPIC can also ensure E2E authenticity since the verification process done by the utility in \autoref{sub:utility_operations} requires the use of symmetric keys shared between the utility and each legitimate user in the AMI network.
Therefore, successful verification process means that the received aggregated reading was computed from the intended system users.

\subsubsection{Key agreement}
%\subsubsection{Attacks on long-term key agreement}
%It is crucial to ensure the security of the proposed key agreement procedures because these keys are used to compute the secret masks that are used to preserve consumer's privacy and defend against collusion attacks.

\textit{Long-term key agreement}.
As shown in Fig. \ref{fig:KeyAgreement}, the two parties that run the long-term key agreement procedure should contribute by random numbers which usually results in more secure key comparing to computing the key by one party.
The security of the key computation relies on the hardness of the Discrete-Logarithmic Problem (DLP).
If $\mathcal{A}$ eavesdrops on the communication between SM$_i$ and $\mathcal{P}_{i,j}$ given in Fig. \ref{fig:KeyAgreement}, he can obtain $v_{i,j}P$ and $v_{j,i}P$.
However, given $v_{i,j}P$ and $P$, it is computationally infeasible to obtain $v_{i,j}$.
Therefore, only the involved parties can compute the keys.

%For Man-In-The-Middle attack (MITM), Assume that $\mathcal{A}$ resides between two meters that need to establish a long-term key. $\mathcal{A}$ may use the messages sent by each meter to fool the other meter and share a key with each meter instead of establishing end-to-end keys shared by the two meters. However, all packets in the long-term key agreement procedure in EPIC are signed to thwart MITM attacks by authenticating the packets' sender.

%For backward and forward secrecy, fresh random numbers are used to compute new keys and the old long-term key is not used in the computation of the new key, and thus even if an attacker was able to compute a long-term key, he cannot use the key to compute past or future keys.

%\subsubsection{Forward and backward secrecy for short-term keys}
\textit{Short-term key agreement}
%If a session key is compromised, attackers can not compute neither previous session keys nor future session keys.
For backward and forward secrecy, as shown in Fig. \ref{fig:SessionKey}, given the current short-term key, $\mathcal{A}$ can compute neither the past keys nor the future keys.
The only way to compute a correct short-term key is to have the two corresponding hash values from the forward and backward chains.
Assuming an attacker could obtain a short-term key $K_{i,j}^{(s)}=H_1(F_{s}\oplus B_{s})$, it is computationally infeasible to extract $F_{s}\oplus B_{s}$ from $K_{i,j}^{(s)}$ because the hash function is irreversible.
Moreover, if $F_{s}\oplus B_{s}$ is given to $\mathcal{A}$, it is infeasible to separate them because they are XORed.
%The next statment can be removed if we want to shorten the paper
Furthermore, assuming $\mathcal{A}$ is given an element from the forward chain $F_{t_x}$ and the corresponding element from the backward chain $B_{t_x}$, since $F_{t_x}$ is hashed forwardly and $B_{t_x}$ is hashed backwardly,
$\mathcal{A}$ can compute neither the previous short-term keys due to the missing forward chain elements nor the
future keys due to the missing backward chain elements.

\begin{table}
	\caption{Computational times and sizes for cryptographic operations.}
	\vspace{-3mm}
	\label{tab:operations}
	\center%
	\begin{tabular}{ccc}
		\toprule
		& Cryptographic Operation  & Time\tabularnewline
		\midrule
		$T_{1}$ & Pairing $e(P_{1},P_{2})$  & 1.025 ms\tabularnewline
		\\[-1.1em]
		$T_{2}$ & $g_{1}\times g_{2}\in\mathbb{G}_{2}\ (512 bit)$ & 1.22 $\mu$s\tabularnewline
		\\[-1.1em]
		$T_{3}$ & $\sigma_{1}+\sigma_{2}\in\mathbb{G}_{1}\ (512 bit)$ & 4.4 $\mu$s\tabularnewline
		\\[-1.1em]
		$T_{4}$ & $H_{2}:\left\{ 0,1\right\} ^{*}\rightarrow\mathbb{G}_{1}$ & 0.05 ms\tabularnewline
		\\[-1.1em]
		$T_{5}$ & $xP_{1}$$\in\mathbb{G}_{1}$ & 1.44 ms\tabularnewline
		\\[-1.1em]
		$T_{6}$ & $\mathcal{H}(m)$$\in\mathbb{G}$ & 1.3 ms\tabularnewline
		\\[-1.1em]
		$T_{7}$ & $h_{1}+h_{2}$$\in\mathbb{G}$ & 1.31 $\mu$s\tabularnewline
		\\[-1.1em]
		$T_{8}$ & HMAC {[}RFC 2104{]} \ (128 bit) & 1.58 $\mu$s\tabularnewline
		\\[-1.1em]
		\midrule
		$T_{9}$ & Modular Exponentiation\ (1024bit) & 5.88 ms\tabularnewline
		\\[-1.1em]
		$T_{10}$ & Point Multiplication\ (1024bit) & 1.36 $\mu$s\tabularnewline
		\\[-1.1em]
		$T_{11}$ & Paillier encryption\ ($1024 bits$) & 19.47 ms\tabularnewline
		\\[-1.1em]
		$T_{12}$ & Paillier decryption\ ($1024 bits$) & 18.88 ms\tabularnewline
		\\[-1.1em]
		$T_{13}$ & Paillier aggregation\ ($1024 bits$) & 19.9 $\mu$s\tabularnewline
		\\[-1.1em]
		$T_{14}$ & Paillier Ciphertext Exponentiation\  & 24.25 ms\tabularnewline
		\bottomrule
	\end{tabular}
	\vspace{-5mm}
\end{table}

\section{Performance Evaluation} \label{sec:Performance-Evaluation}

In this section, we first evaluate EPIC in terms of the communication and computation overheads for the single-hop model, then, we present our ns-3 experiment results to assess the network performance for the single- and multi-hop models.

\subsection{Computation and Communication Overhead}

\begin{figure*}[!t]
	\setlength{\abovecaptionskip}{0.1cm} % 0.5cm as an example
	\setlength{\belowcaptionskip}{-0.7cm} % 0.5cm as an example
	\centering
%\subfloat[SM computations.\label{fig:sm_computations}] {\includegraphics[width=0.29\linewidth]{sm_computations.eps}}
%\subfloat[Gateway computations.\label{fig:gw_computations}] {\includegraphics[width=0.29\linewidth]{gw_computations.eps}}
%\subfloat[Utility Computations.\label{fig:u_computations}] {\includegraphics[width=0.29\linewidth]{u_computations.eps}}
	\subfloat[SM computations.\label{fig:sm_computations}]
    {\includegraphics[width=0.3333333333\linewidth]{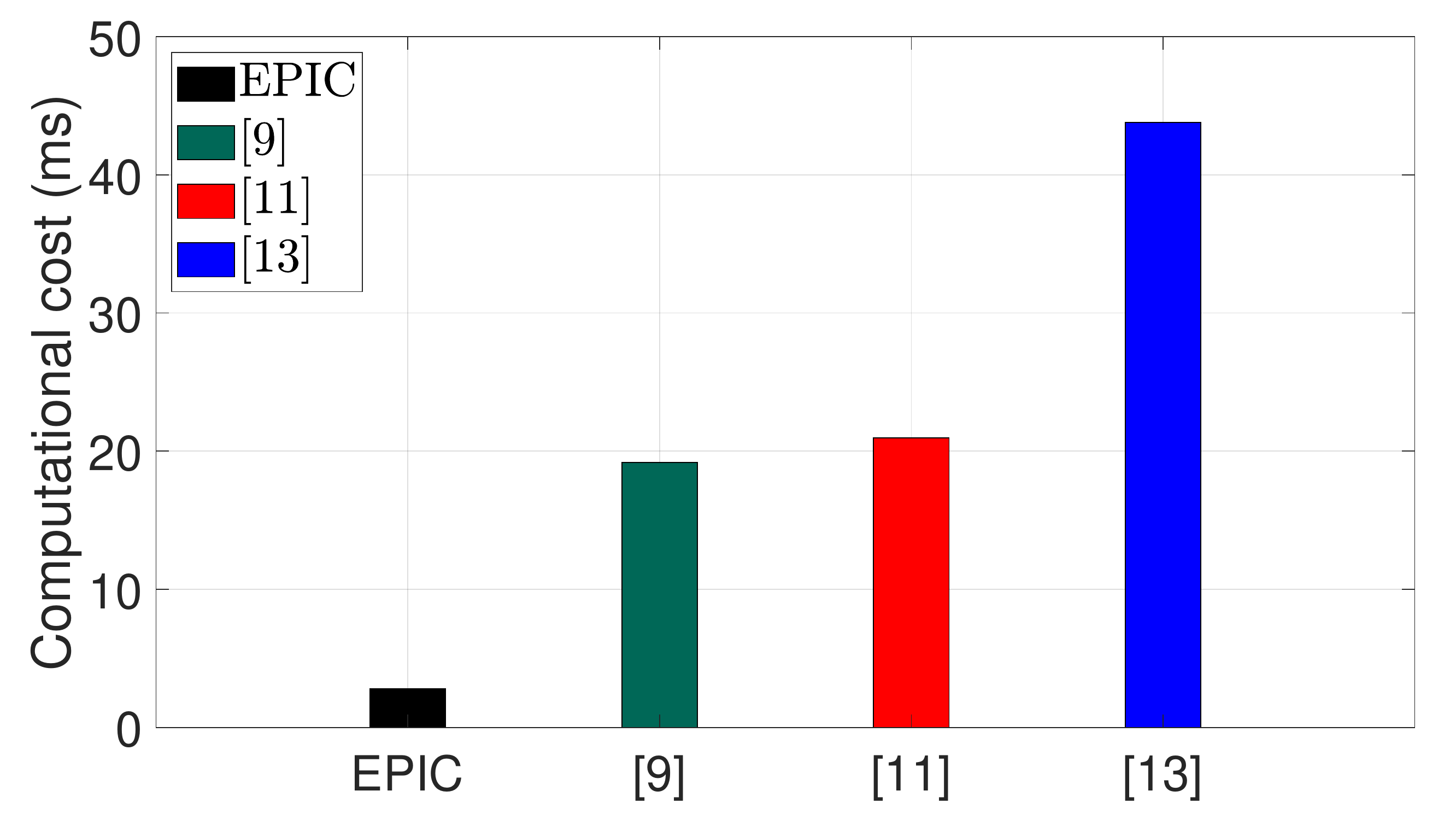}}
    \subfloat[Gateway computations.\label{fig:gw_computations}]
	{\includegraphics[width=0.3333333333\linewidth]{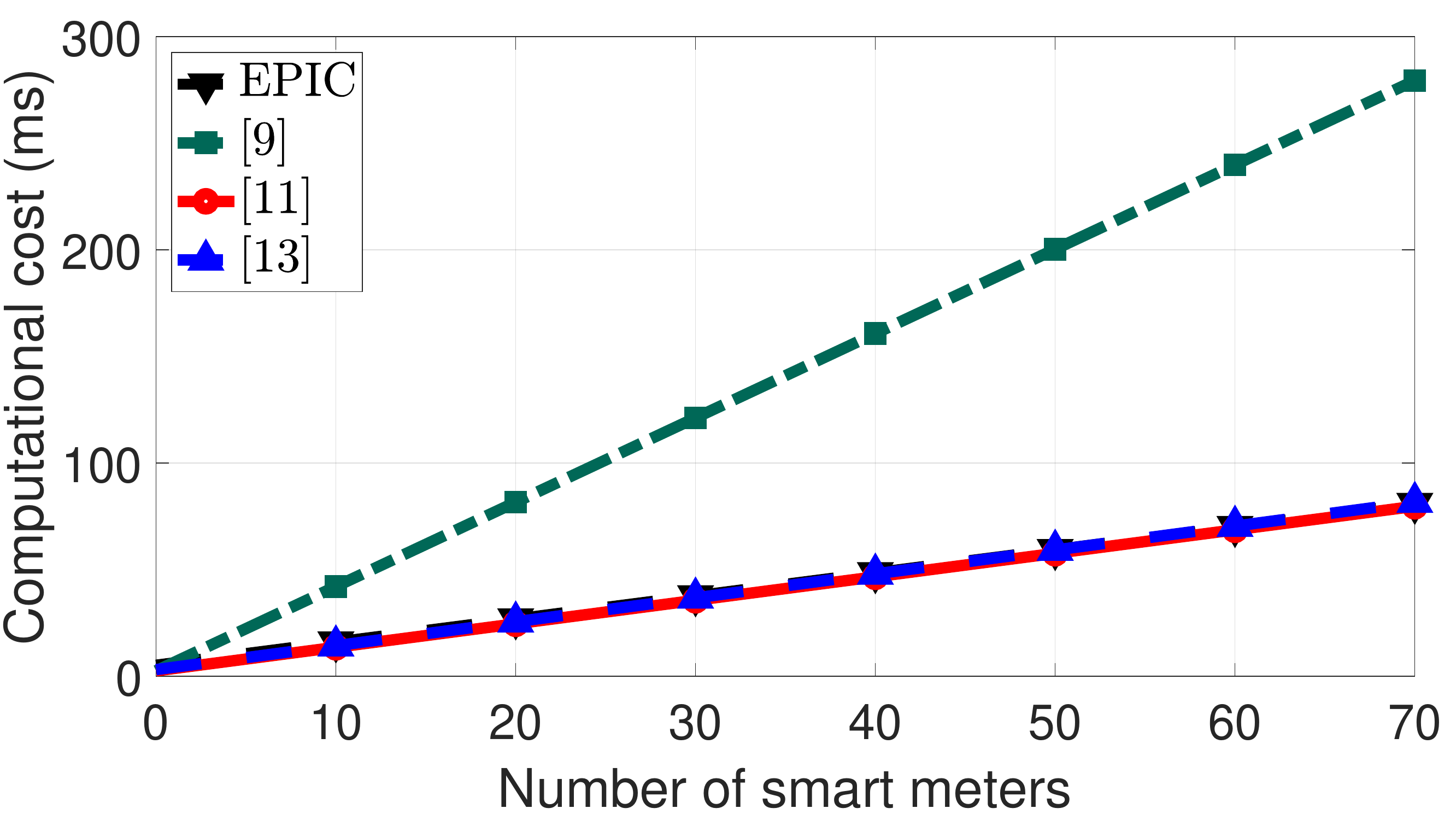}}
    \subfloat[Utility Computations.\label{fig:u_computations}]
	{\includegraphics[width=0.3333333333\linewidth]{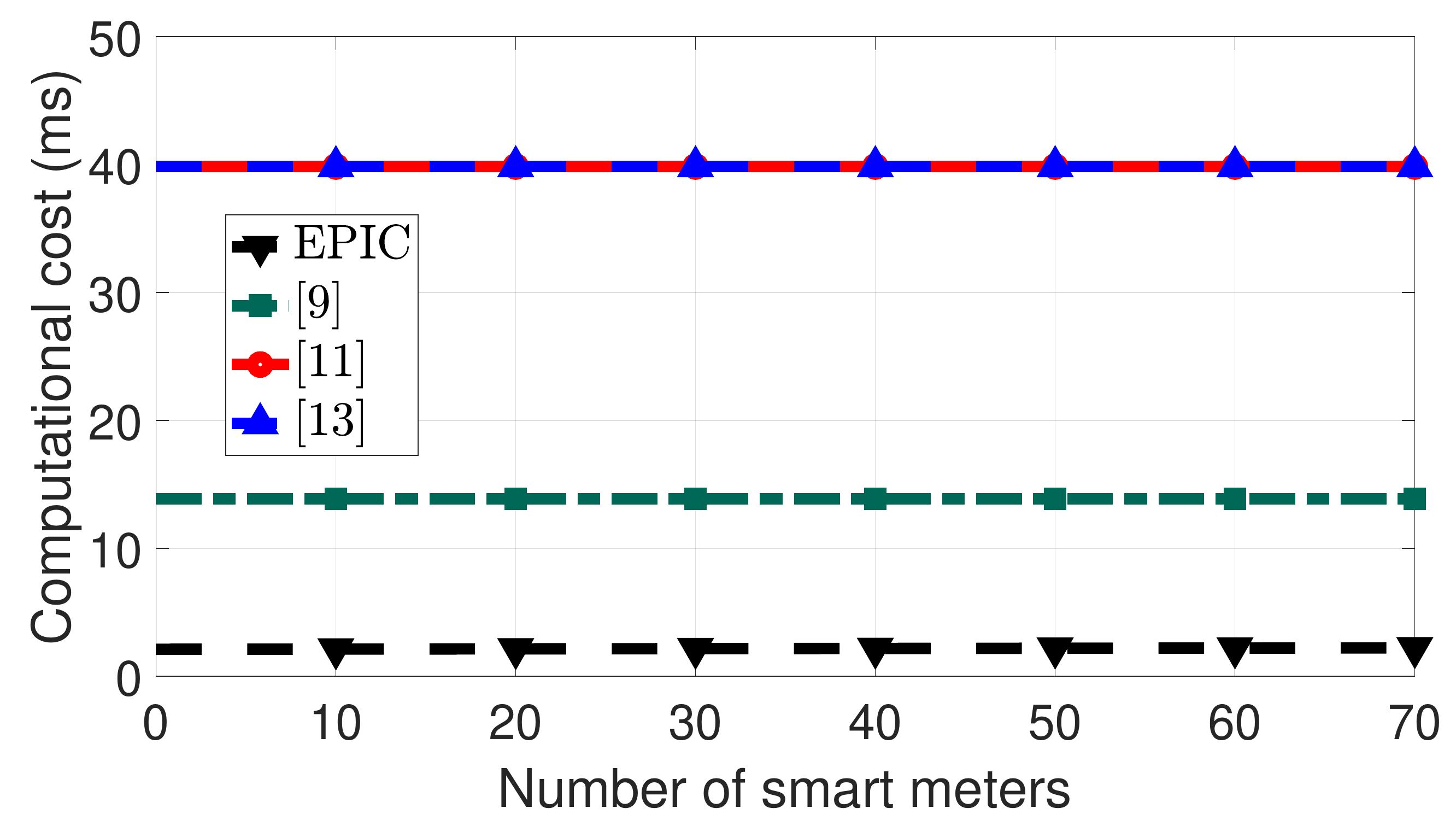}}
	
	\caption{Computation overhead comparison}
	\label{fig:computation_new}
	\vspace{-5mm}
\end{figure*}

%\begin{figure*}[!t]
%	\setlength{\abovecaptionskip}{0.1cm} % 0.5cm as an example
%	\setlength{\belowcaptionskip}{-0.7cm} % 0.5cm as an example
%	\centering
%	\subfloat[SM-to-Gateway overhead.\label{fig:sm_gw_overhead}]
%	{\includegraphics[width=0.3\linewidth]{sm_gw_overhead.eps}}
%	\subfloat[Gateway-to-Utility overhead. \label{fig:gw_u_overhead}]
%	{\includegraphics[width=0.3\linewidth]{gw_u_overhead.eps}}
%	
%	\caption{Communication overhead comparison.}
%	\label{fig:overhead_new}
%		\vspace{-5mm}
%\end{figure*}

%\begin{figure}[!t]
%	
%	\centering
%	\subfloat[SM-to-Gateway overhead. \label{fig:sm_gw_overhead}]
%	{\includegraphics[scale=0.55]{sm_gw_overhead.eps}}
%	
%	\subfloat[Gateway-to-Utility overhead. \label{fig:gw_u_overhead}]
%	{\includegraphics[scale=0.55]{gw_u_overhead.eps}}
%	
%	\caption{Communication overhead comparison.}
%	\label{fig:overhead_new}
%	\vspace{-3mm}
%\end{figure}

To evaluate the communication and computation overheads of EPIC, we implemeted the required cryptographic operations using Python charm cryptographic library \cite{charm} running on an Intel Core i7-4765T 2.00 GHz and 8 GB RAM.
We used supersingular elliptic curve with the symmetric Type 1 pairing of size 512 bits (SS512 curve) for bilinear pairing and a standard elliptic curve secp160r1 for the homomorphic hash function \cite{Certicom_Corp}.
%Table \ref{tab:Sizes} summarizes the sizes of each cryptographic component used to construct the exchanged messages in EPIC and the schemes under comparison.
%In addition, all cryptographic operations were run 1000 times and average measurements are reported in Table \ref{tab:operations}.
All cryptographic operations were run 1,000 times and average measurements are reported in the upper part of Table \ref{tab:operations}.
Since we compare the overhead of EPIC to the proposed schemes in \cite{HT20,cube,DI01}, we include in the lower part of Table \ref{tab:operations} the computation measurements of
the cryptographic operations needed in these schemes.
%We compare our scheme to Paillier cryptosystem based schemes.
%In addition, we added the same security features achieved in our scheme to Paillier cryptosystem based schemes \cite{HT20,6165271,DI01}.

%\begin{table}
%	\center%
%	\caption{Size of each cryptographic component of a SM$_i$ message. \label{tab:Sizes}}
%	\begin{tabular}{cc}
%		\toprule
%		Cryptographic Component & Size (Bytes)\tabularnewline
%		\midrule
%		$M_{i}$ & 16\tabularnewline
%		\\[-0.9em]
%		$TS$ & 4\tabularnewline
%		\\[-0.9em]
%		$h_{i}\in\mathbb{G}$ & 20\tabularnewline
%		\\[-0.9em]
%		$MAC_i$ & 16\tabularnewline
%		\\[-0.9em]
%		$\sigma_{i}\in\mathbb{G}_1$  & 64\tabularnewline
%		\\[-0.9em]
%		%$E_{AES}(.)$ & 16/Block\tabularnewline
%		%\\[-0.9em]
%		$E_{Paillier}(.)$ & 512\tabularnewline
%		\bottomrule
%	\end{tabular}
%\end{table}

\subsubsection{Computation Overhead}

%For SMs, the computation overhead is measured by the time required to construct a packet.
%For the gateway, it is time required to verify the received packets from SMs plus the time required to construct a packet.
%For the utility, it is time required to verify the received packet from the gateway plus the time required to decrypt the received ciphertext to obtain the aggregated reading.
The computation overhead is defined as the processing time required by each node in the network. These nodes are SMs, the gateway and the utility.

For the single hop model, the time-consuming operations required by SMs are one homomorphic hash generation which requires $T_{6}$;
one HMAC generation which requires $T_{8}$; and one signature generation which requires $T_4+T_5$.
Using the measurements in Table \ref{tab:operations}, the total time required by each meter is 2.79 ms.
For $n$ SMs, the computations required by the gateway are
batch signatures verification (as in Eq. \ref{eq:sign}) which requires $(n+1)T_{1}+(n-1)T_{2}+(n-1)T_{3}+nT_{4}$;
batch homomorphic hashes verification (as in Eq. \ref{eq:verify}) which requires $T_{6}+(n-1)T_{7}$ ;
one homomorphic hash generation which requires $T_{6}$;
one HMAC generation which requires $T_{8}$;
%one AES encryption which requires $\ceil{\frac{20+20\ell_{i}}{16}}T_{9}$;
and one signature generation which requires $T_4+T_5$.
The total time required by the gateway for these operations is $1.0836n+5.1128$ ms.
For the utility, one signature verification operation plus $n$ HMAC computations are required to verify the received packet and one arithmetic addition operation to obtain the aggregated reading.
These operations require $0.001n+2.1037$ ms.
For the schemes in \cite{HT20,cube,DI01}, we followed the same procedure to compute the computation overhead
of each entity and the results are presented in Fig. \ref{fig:computation_new}.

As shown in Fig. \ref{fig:computation_new}(a), EPIC imposes the least computation overhead on the SMs comparing to the existing schemes.
This is because EPIC uses efficient masking technique, while the other schemes use computationally-extensive operations to encrypt the fine-grained readings.
For the gateway, Fig. \ref{fig:computation_new}(b) shows that the computational overhead of the gateway in EPIC is found to be close to those of \cite{cube} and \cite{DI01}.
%since these schemes uses BLS signature verification process.
For the computation overhead of the utility, EPIC is more efficient than the existing schemes because simple arithmetic addition is needed to remove the utility mask and recover the aggregated reading as shown in equation \ref{eq:agg_read_final} , while in the schemes of \cite{HT20,cube,DI01}, time-consuming decryption operation is needed, as given in the lower part of \autoref{tab:operations}.
In addition, as shown in Fig. \ref{fig:computation_new}(c), the proposed schemes in \cite{HT20,cube,DI01} have constant computation time because the utility decrypts only one aggregated ciphertext regardless of the number of SMs, whereas in EPIC, the utility's computation overhead increases linearly at a rate of 1.31 $\mu$s/SM because the utility receives one homomorphic hash for every SM, and thus more operations are needed.
However, the utility's computation overhead of EPIC is much less than those of the other schemes.\\

%Comparing to Paillier cryptosystem scheme, our scheme has lower computation overhead.
%This is because the time needed to encrypt a reading using Paillier cryptosystem ($T_{11}$) is much higher than the time required to mask the reading in our scheme which is a simple arithmetic addition.
%The measurements presented in this subsection are used in the ns-3 experiments discussed in \autoref{sub:ns3}.

% ===========  ns3 Figures  ===========
\begin{figure*}[!t]
	\setlength{\abovecaptionskip}{0.1cm} % 0.5cm as an example
	\setlength{\belowcaptionskip}{-0.7cm} % 0.5cm as an example
	\centering
	
	\subfloat[Completion time for E2E model.\label{fig:E2ECT}]
	{\includegraphics[width=0.3333333333\linewidth]{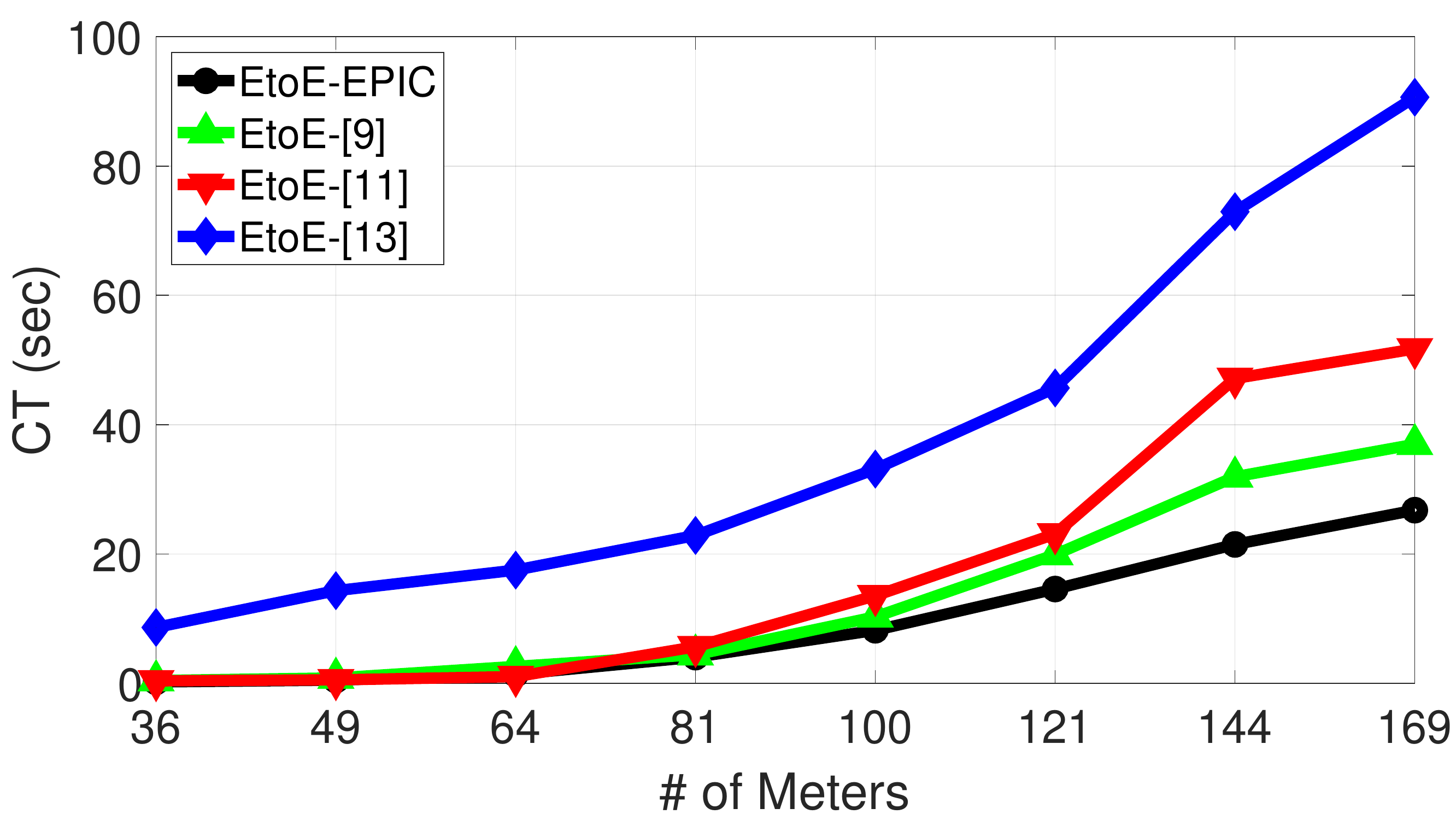}}	
	\subfloat[Throuhput for E2E model.\label{fig:E2ETP}]
	{\includegraphics[width=0.3333333333\linewidth]{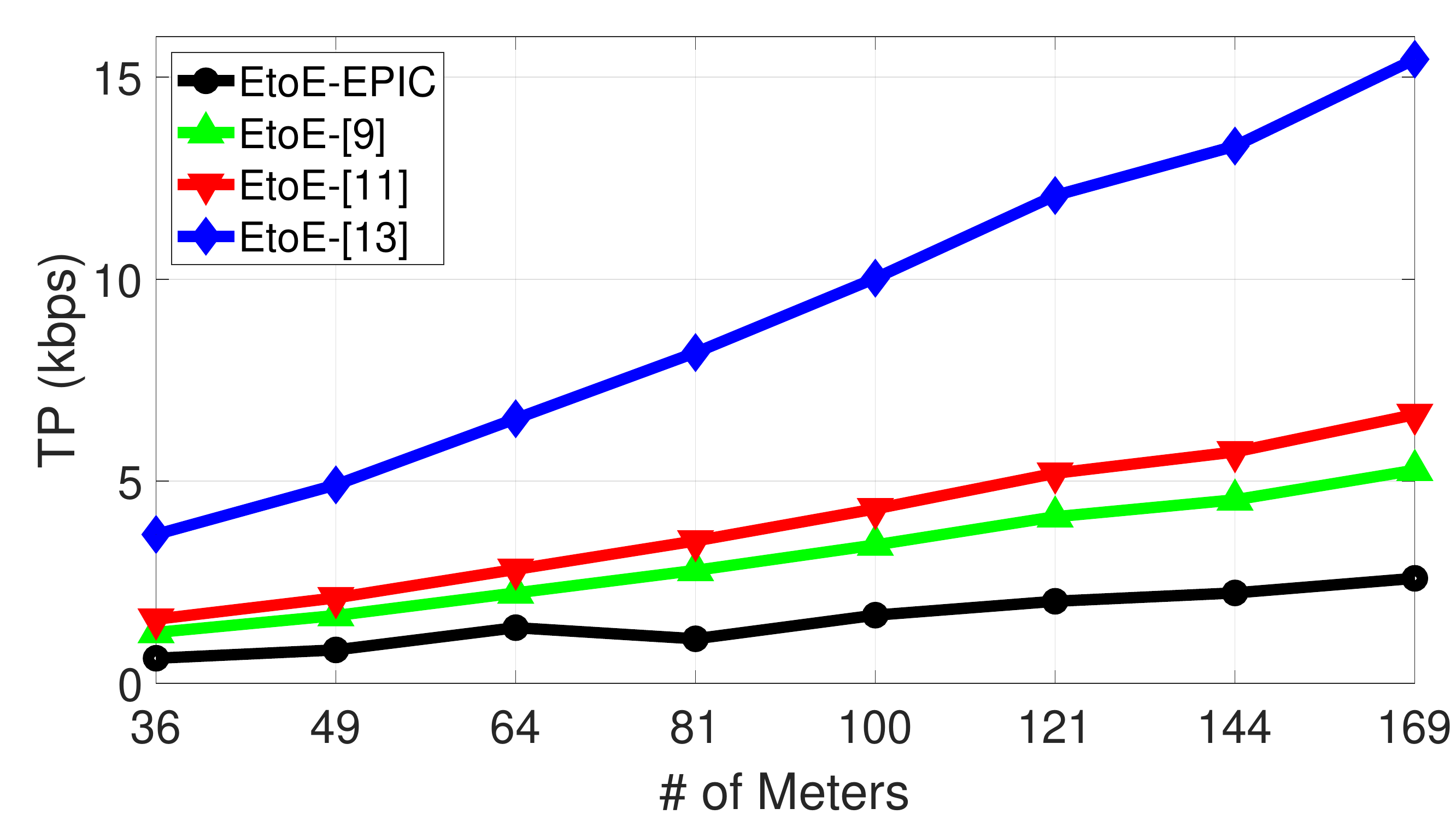}}	
	\subfloat[Packet delivery ratio for E2E model.\label{fig:E2EPDR}]
	{\includegraphics[width=0.3333333333\linewidth]{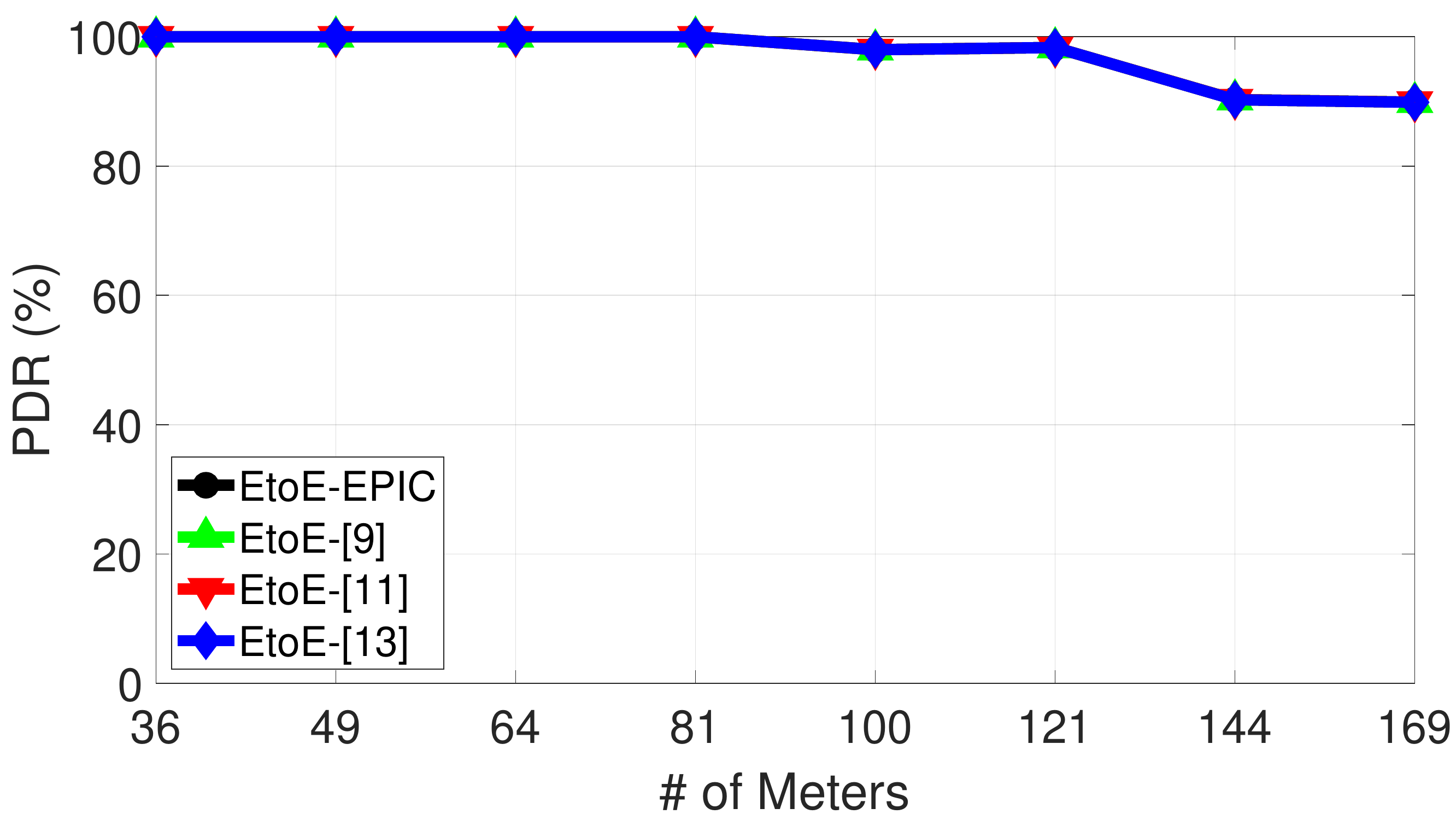}}\\
	\vspace{-3 mm}
	\subfloat[Completion time  for HbyH model.\label{fig:HbHCT}]
	{\includegraphics[width=0.3333333333\linewidth]{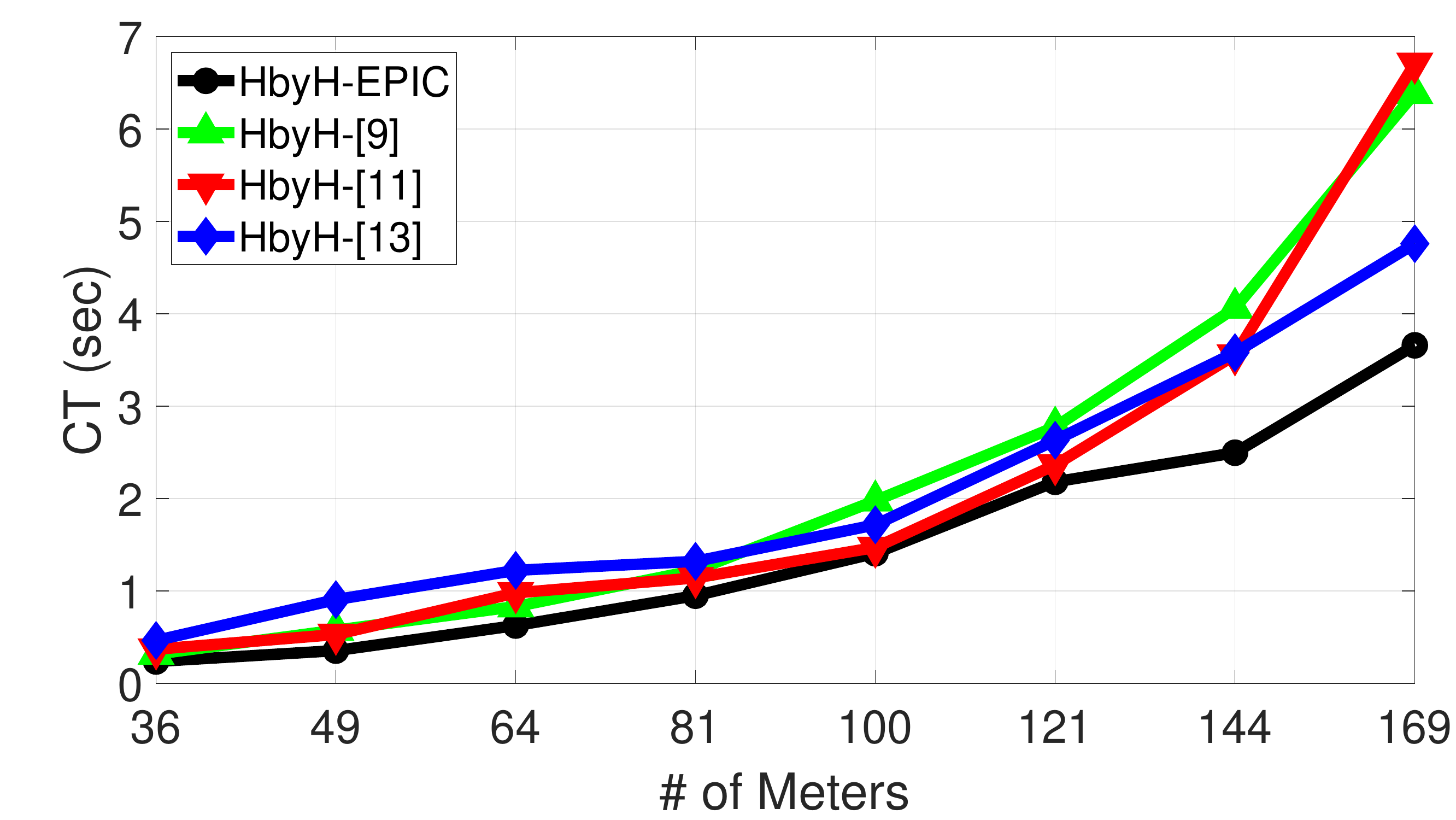}}	
	\subfloat[Throuhput for HbyH model.\label{fig:HbHTP}]
	{\includegraphics[width=0.3333333333\linewidth]{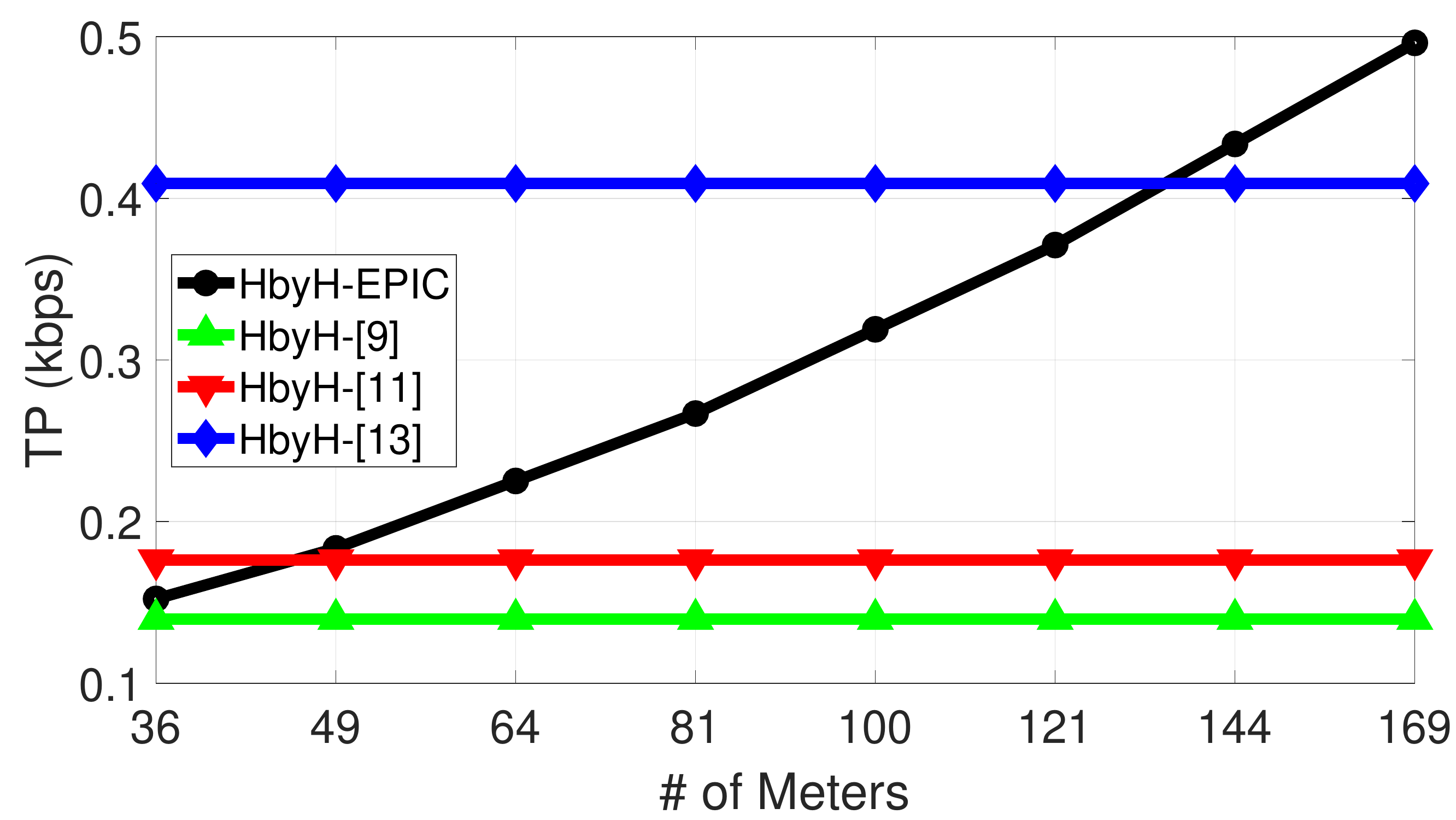}}	
	\subfloat[Packet delivery ratio for HbyH model.\label{fig:HbHPDR}]
	{\includegraphics[width=0.3333333333\linewidth]{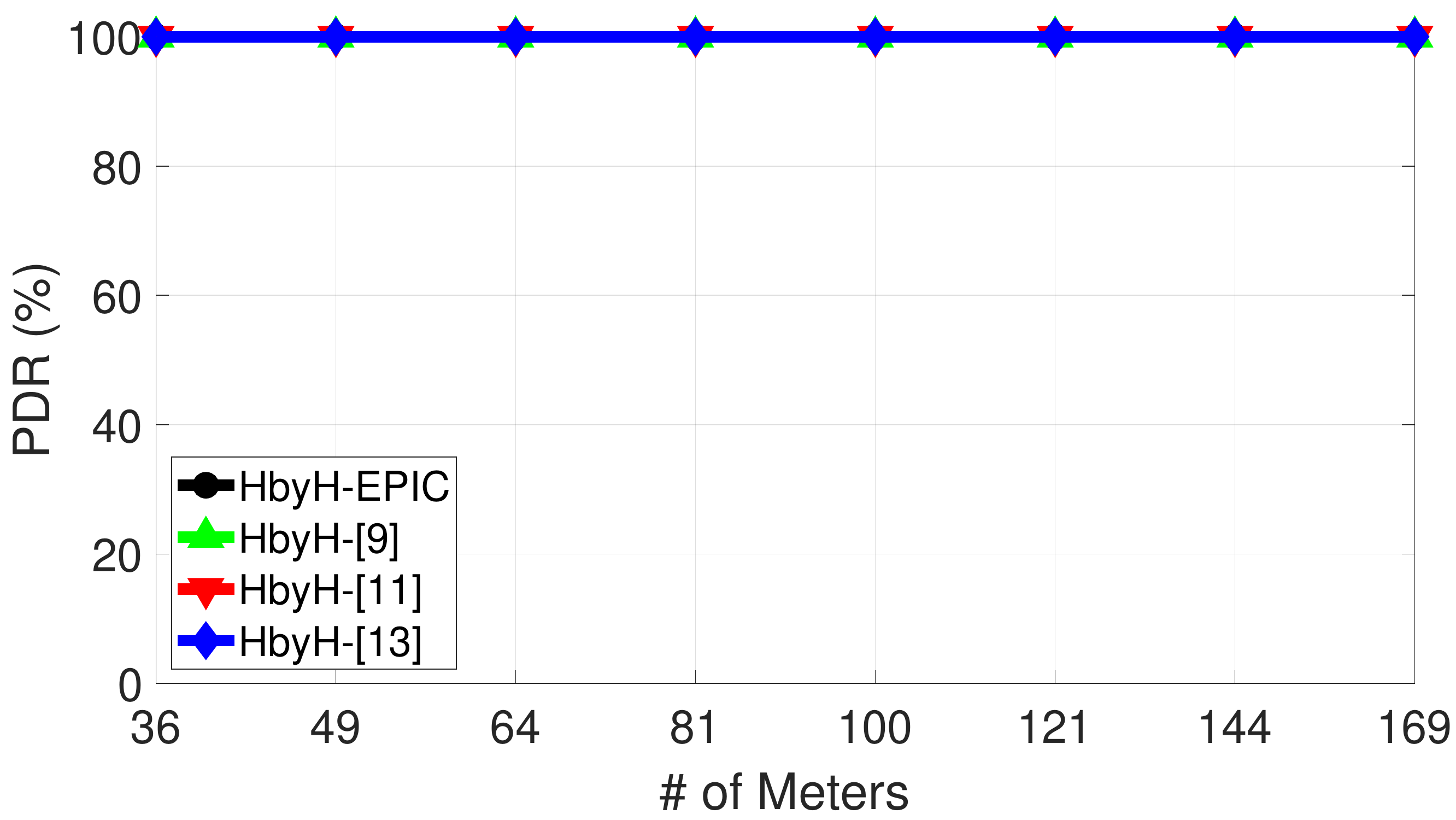}}	
	\caption{Computation overhead comparison}
	\label{fig:ns3_new}
	\vspace{-5mm}
\end{figure*}

\subsubsection{Communication Overhead}

% Before removing the overhead table
%The communication overhead can be measured by the size, in bytes, of transmitted messages between the network entities. In sepceific, we have two overheads to be evaluated, SM-to-Gateway overhead and Gateway-to-Utility overhead.
%
%The SM-to-Gateway overhead in EPIC can be computed based on equation \ref{eq:leaf_report} and Table \ref{tab:Sizes} as follows. Each SM transmits a 16-byte masked reading, a 4-byte timestamp, a 20-byte homomorphic hashe, a 16-byte MAC and a 64-byte signature. Therefore, the total size of a SM's message is 120 bytes.
%On the other hand, the Gateway-to-Utility overhead depends mainly on $n$ which is the total number of SMs in the network.
%This is because the gateway should forward to the utility all the individual homomorphic hasesh.
%Therefore, the total size of the gateway message is $20*n100$ bytes.
%We followed the same procedure to obtain the communication overhead for \cite{HT20}, \cite{6165271} and \cite{DI01} and the comparison is presented in \hl{figure overhead}.

% After removing the overhead table
The communication overhead is measured by the size of transmitted messages between the network entities in bytes.
In specific, we evaluate SM-to-Gateway and Gateway-to-Utility communication overhead.

The SM-to-Gateway communication overhead in EPIC can be computed using the packet format in (\ref{eq:leaf_report}) as follows.
Each SM transmits a 16-byte masked reading, a 4-byte timestamp, a 20-byte homomorphic hashe, a 16-byte MAC and a 64-byte signature.
Therefore, the total size of a SM's message is 120 bytes.
For a network of size $n$, the communication overhead between all SMs and the gateway is $120n$.
On the other hand, the Gateway-to-Utility communication overhead depends mainly on $n$ which is the total number of SMs in the network.
The total size of the gateway message to the utility is $20n+100$ bytes.
We used the same procedure to compute the communication overhead of the proposed schemes in \cite{HT20}, \cite{cube} and \cite{DI01}.

The values of computations and communication overhead presented in this section were used within the ns-3 simulation presented in the following subsection.
% , and the communication overheads of these schemes are compared to the overhead of EPIC in Figs. \ref{fig:sm_gw_overhead} and \ref{fig:gw_u_overhead}.

% Removed after ns3 is added
%Fig. \ref{fig:sm_gw_overhead} shows that EPIC has the lowest SM-to-Gateway communication overhead.
%This is because the size of the masked reading in EPIC is much smaller than the ciphertext size in the other schemes.
%Fig. \ref{fig:gw_u_overhead} shows that the Gateway-to-Utility overhead in the proposed schemes in \cite{HT20}, \cite{6165271}, and \cite{DI01} are constant while it increase linear in EPIC.
%This is because the gateway should forward to the utility all the individual homomorphic hashes which are needed to achieve E2E data integrity verification and compute the dynamic-pricing-based bills
%that are not achieved in the other schemes.

%\begin{figure}[!t]
%
%	\centering
%	\subfloat[SM-to-Gateway overhead. \label{fig:sm_gw_overhead}]
%	{\includegraphics[scale=0.55]{sm_gw_overhead.eps}}
%	
%	\subfloat[Gateway-to-Utility overhead. \label{fig:gw_u_overhead}]
%	{\includegraphics[scale=0.55]{gw_u_overhead.eps}}
%	
%	\caption{Communication overhead comparison.}
%	\label{fig:overhead_new}
%	\vspace{-3mm}
%\end{figure}

\subsection{Experiment and Measurement Results} \label{sub:ns3}

%\hl{This subsection is from the conference version and should be updated after new simulation}

\subsubsection{Experimental Setup}

%\begin{figure*}[!t]
%	\setlength{\abovecaptionskip}{0.1cm} % 0.5cm as an example
%	\setlength{\belowcaptionskip}{-0.7cm} % 0.5cm as an example
%	\centering
%	\subfloat[Average completion time.\label{fig:CT}]
%	{\includegraphics[width=0.333\linewidth]{CT.pdf}}
%	\subfloat[Throughput as bandwidth usage indicator.\label{fig:TP}]
%	{\includegraphics[width=0.333\linewidth]{TP.pdf}}
%	\subfloat[PDR values as a reliability indicator.\label{fig:PDR}]
%	{\includegraphics[width=0.333\linewidth]{PDR.pdf}}
%	
%	\caption{Computation overhead comparison}
%	\label{fig:ns3_new}
%	\vspace{-5mm}
%\end{figure*}

We used network simulator ns-3.27 \cite{ns327} to assess the impact of the communication/computation overhead reduction on the network performance.
We implemented a wireless mesh network that mimics the AMI network using IEEE 802.11s standard.
The underlying MAC protocol is IEEE 802.11g.
TCP was used at the transport layer for a reliable communication, and maximum segment size (MSS) is 536 bytes.

We created grid topologies of size \textbf{$N$}, where \textbf{$N$} $\in \{36, 49, 64, 81, 100, 121, 144, 169\}$. For each \textbf{$N$}, we ran the schemes for 30 rounds and average results are reported.
One of the nodes in each topology acts as the gateway while the other (\textbf{$N-1$}) nodes act as smart meters.

At the beginning of each data collection cycle, each meter reports its power consumption reading to the gateway.
We assumed that the data collection is performed periodically every 60 seconds \cite{beussink2014preserving}.
We simulated two different network models: End-to-End (EtoE) and Hop-by-Hop (HbyH).
In EtoE data collection model, all smart meters simultaneously report their readings to the gateway directly and multi-hop packet relay may be needed.
In HbyH model, a minimum spanning tree of the network is created, and parent-child relationships are assigned.% based on this tree.
Moreover, in HbyH model, leaf meters send their readings to their parent meter periodically, the parent meter aggregates its reading with the readings received from the child meters, and sends an aggregated reading to their parent meter.
This process goes on up to the gateway.
Finally, the gateway aggregates the readings received from its child meters and sends an aggregated reading to the utility.

\subsubsection{Baselines and Performance Metrics}

We use three existing works \cite{HT20,cube,DI01} as baselines to compare the performance of EPIC.
In \cite{HT20}, the meters send their readings in blinded form, and the data aggregation is performed on ciphertext. The total power usage can be recovered by computing a discrete log problem.
The scheme in \cite{cube} uses Paillier cryptosystem to perform data aggregation using partially homomorphic encryption. %This baseline provides both security and privacy since it is able to perform arithmetic addition on encrypted data. However, it generates larger data packets when compared to \cite{HT20}.
This baseline generates larger data packets when compared to \cite{HT20}.
The last baseline \cite{DI01} also uses Paillier cryptosystem, but it differs from \cite{cube} in that it uses two signatures for each report. Hence, \cite{DI01} introduces extra overhead.

For performance evaluation, we used the following metrics:

\begin{itemize}
	\item \textit{Average Completion Time (CT)}:
	It is the elapsed time for gathering all the measurement data from all of the nodes and aggregating them at the gateway in one cycle.
	We measure CT at the application layer so that the cryptographic operations are taken into account.
	\item \textit{Throughput (TP)}:
	It is the average amount of data received by the gateway per second. This parameter is an indicator for the bandwidth usage of each scheme, i.e., as measurement of this metric increases, as it is worse.
	\item \textit{Packet Delivery Ratio (PDR)}:
	It is the ratio of the number of data packets received by the gateway to the number of data packets that are expected to be received by the gateway.
\end{itemize}

\subsubsection{Simulation Results and Discussions}

%\begin{figure}[!t]
%	\center
%	\includegraphics[scale=0.32]{CT1.pdf}
%	%\vspace{-1mm}
%	\caption{Average completion time.}
%	\label{fig:CT}
%	%\vspace{-3mm}
%\end{figure}

%In Fig. \ref{fig:ns3_new}, we present the data collection completion time values.
%It can be seen that as the network grows, the time required to complete a data collection cycle increases.
%In the HbyH model, CT increases slowly while it dramatically increases in the EtoE model especially in case of 100 nodes.
%This difference can be attributed to the propagation delay mostly.
%It includes backoff waitings due to external collisions while accessing the medium to transmit the data packets and the path discovery process performed by the HWMP which is the default routing protocol of IEEE 802.11s standard \cite{bahr2007update}.
%It can also be seen that the CT in EPIC is very close to that of the plaintext scheme.

In Fig. \ref{fig:ns3_new}(a) and Fig. \ref{fig:ns3_new}(d), we present the data collection completion time values.
As the network grows, the time required to complete a data collection cycle increases.
In both data collection methods, EPIC requires the least time for all topology sizes in a round because it both has a moderate processing delay for aggregation and generates comparable size of power readings.
In addition, the approaches require similar amount of time for data collection until 81-node topologies.
Thereafter, the values dramatically increase for EtoE data collection.
This difference can be attributed to the propagation delay mostly.
It includes backoff waitings due to external collisions while accessing the medium to transmit the data packets and the path discovery process performed by the HWMP which is default routing protocol of IEEE 802.11s standard~\cite{bahr2007update}.
The EtoE data collection typically needs more hops to deliver data packets to the destination because parent and child meters are one-hop neighbors of each other in the data reporting hierarchy trees.
Thus, the data packets are exposed to more backoff waitings on the path towards the destination.
This results in a dramatic increase in EtoE data collection.
Also, we would like to point out the remarkable difference between \cite{DI01} and the other approaches.
Although \cite{DI01} takes shorter than \cite{HT20} does to perform cryptographic operations, it incurs an extra delay due to the segmentation by TCP at the transport layer.
Since \cite{DI01} generates larger data packets than MSS, a power reading is transmitted in multiple segments, which results in an extra delay due to the extra backoff waitings.

It can be seen that in HbyH data collection that the approaches require far less time to complete a data collection round. Also, they have similar values for all topology sizes. Since parent and child meters are one-hop neighbor of each other, the backoff waitings decrease thanks to less competition in medium access.
EPIC slightly outperforms the other approaches. It always requires less time than the other approaches do to complete a data collection round. Although it incurs similar processing delay for aggregation and generates larger aggregated readings beyond a level of the hierarchy tree towards the gateway, it requires the least time. This is because EPIC takes advantage of far shorter propagation delays below that level thanks to smaller aggregated power readings \cite{korhonen2005effect}.

%\begin{figure}[!t]
%	\center
%	\includegraphics[scale=0.32]{TP1.pdf}
%	%\vspace{-1mm}
%	%\caption{The throughput as an indicator of bandwidth usage.}
%	\caption{Throughput as bandwidth usage indicator.}
%	\label{fig:TP}
%	%\vspace{-3mm}
%\end{figure}

%Old for comparison against plain text
%Fig. \ref{fig:TP} gives the throughput (TP) versus the number of meters in the network.
%As shown in the figure, all the approaches produce more throughput at the gateway in the EtoE model when compared to the TP values of the HbyH model.
%This is because power readings are aggregated at some intermediate meters in the HbyH model.
%Hence, the average amount of data received by the gateway decreases.
%In addition, EPIC produces more throughput in both the EtoE and HbyH models since it generates the largest data packets compared to the plaintext.
%The additional overhead is required to achieve E2E data integrity and dynamic pricing.

Secondly, we analyze the throughput performance to discuss the bandwidth usage of the approaches.
As shown in  Fig. \ref{fig:ns3_new}(b) and Fig. \ref{fig:ns3_new}(e), the approaches produce more throughput at the gateway in the EtoE data collection when compared to the throughput values for the HbyH data collection.
This is because power readings are aggregated at some intermediate meters in the HbyH model.
Hence, average amount of data received by the gateway decreases.
\cite{DI01} produces the most throughput in both the EtoE and HbyH data collection methods since it generates the biggest data packets compared to the others.
It is followed by \cite{cube} and \cite{HT20}, respectively.
EPIC produces the least throughput because it generates the smallest data packets for power readings.
The TP values linearly increase in EtoE data collection because the number of power readings delivered to the gateway increases as the network grows.
However, in HbyH data collection, the throughput that EPIC produces at the gateway linearly increases while the others remain constant although the number of power readings reported by each approach is the same at each topology size.
%This is because the size of power readings aggregated by a parent meter depends on the number of all meters under its hierarchy when EPIC is used.
The additional overhead is required to achieve E2E data integrity, authenticity  and dynamic pricing that are not achieved in the other schemes.

%\begin{figure}[!t]
%	\center
%	\includegraphics[scale=0.32]{PDR1.pdf}
%	%\vspace{-1mm}
%	\caption{PDR values as a reliability indicator.}
%	\label{fig:PDR}
%	%\vspace{-3mm}
%\end{figure}

We investigate the PDR values in order to find out how reliable the schemes are.
As shown in Fig. \ref{fig:ns3_new}(c) and Fig. \ref{fig:ns3_new}(f), all schemes achieve more than 89\% PDR.
While the schemes can achieve 100\% at each topology size in the HbyH data collection, the values slightly decrease after 81-node topology in the EtoE data collection.
This is due to the loss of one of three-way handshake messages between the gateway and a physically distant node (especially the nodes at the edge of the network) in the network.
The TCP is a connection-oriented communication protocol, so it needs to establish a connection before sending data packets.
As the number of hops between the hosts increases, it is more likely to lose any of the three-way handshake messages.
Since there is a limit to retransmit these messages, it is likely to fail a connection.
If the connection fails, the data packets cannot be transferred, and this results in lower PDR values.

\section{Conclusion} \label{sec:Conclusion}

In this paper, we proposed EPIC, an efficient privacy-preserving scheme with E2E data integrity verification and collusion-resistance for AMI networks.
In EPIC, each meter selects a number of meters in the network called ``\textit{proxies}'' and efficiently computes and shares pairwise secret mask with each proxy.
Then, each meter should  send the utility a masked power consumption reading.
After aggregation, all masks are canceled and the utility can only obtain an aggregated reading to preserve consumers privacy.
In addition, EPIC enables the utility to verify the integrity of the aggregated reading and identify the attackers without accessing the individual readings to preserve privacy.
The utility can also generate electricity bills based on dynamic prices without violating consumers' privacy.
A formal security proof, and a probabilistic model are provided to demonstrate that EPIC can preserve the consumers' privacy with E2E data integrity and high protection against collusion attacks.
Moreover, we evaluated the performance of EPIC using ns-3 and the measurements demonstrated that EPIC is efficient when compared to similar existing schemes and can collect periodic power consumption data in the AMI network without consuming excessive bandwidth so that the other types of traffic can obtain more bandwidth.

%\section{Acknowledgement}
%This work is supported by US National Science Foundation under the grants numbers CNS-1550313 and CNS-1619250.

\bibliographystyle{IEEEtran}
\bibliography{refs_new}

\end{document}